# Scalable synthesis and characterization of multilayer γ-graphyne, new carbon crystals with a small direct bandgap

Authors: Victor G. Desyatkin[1]‡, William B. Martin[1]‡, Ali E. Aliev[2], Nathaniel E. Chapman[1], Alexandre F. Fonseca[3], Douglas S. Galvão[3], E. Roy Miller[4], Kevin H. Stone[5], Zhong Wang[2], Dante Zakhidov[6], F. Ted Limpoco[7], Sarah R. Almahdali[1], Shane M. Parker[4], Ray H. Baughman[2], and Valentin O. Rodionov[1]*

Affiliations:

[1]Department of Macromolecular Science and Engineering, Case Western Reserve University; 2100 Adelbert Road, Cleveland, OH 44106, USA.

[2]Alan G. MacDiarmid NanoTech Institute, University of Texas at Dallas; 800 West Campbell Road, Richardson, TX 75080, USA.

[3]Applied Physics Department, Institute of Physics "Gleb Wataghin", University of Campinas; Campinas, SP, 13083-970, Brazil.

[4]Department of Chemistry, Case Western Reserve University; 10900 Euclid Ave, Cleveland, OH 44106, USA.

[5]Stanford Synchrotron Radiation Lightsource, SLAC National Accelerator Laboratory; 2575 Sand Hill Road, Menlo Park, CA 94025, USA.

[6]Department of Materials Science and Engineering, Stanford University; 496 Lomita Mall, Stanford, CA 94305, USA.

[7]Oxford Instruments Asylum Research; 6310 Hollister Ave, Santa Barbara, CA 93117, USA.

**Abstract:** γ-Graphyne is the most symmetric $sp^2/sp^1$ allotrope of carbon, which can be viewed as graphene uniformly expanded through insertion of two-carbon acetylenic units between all the aromatic rings. To date, synthesis of bulk γ-graphyne has remained a challenge. We here report the synthesis of multilayer γ-graphyne through crystallization-assisted irreversible cross-coupling polymerization. Comprehensive characterization of this new carbon phase is described, including synchrotron X-ray diffraction, electron diffraction, lateral force microscopy, Raman and infrared spectroscopy, and cyclic voltammetry. Experiments indicate that γ-graphyne is a 0.48 eV bandgap semiconductor, with a hexagonal *a*-axis spacing of 6.88 Å and an interlayer spacing of 3.48 Å, which is consistent with theoretical predictions. The observed crystal structure has an aperiodic sheet stacking. The material is thermally stable up to 240 °C but undergoes a transformation at higher temperatures. While conventional 2D polymerizations and reticular chemistry rely on error correction through reversibility, we demonstrate that a periodic covalent lattice can be synthesized under purely kinetic control. The reported methodology is scalable and inspires extension to other allotropes of the graphyne family.

**Introduction:** Over five hundred carbon phases have been theoretically predicted, but few have been experimentally realized.[1] Allotropes based on hexagonal lattices of sp$^2$ hybridized carbons, including few-layer graphene,[2] nanotubes,[3] and fullerenes,[4] are synthetically accessible at scale. These materials have been revolutionary for fundamental physics and materials science, and found applications in post-silicon electronics, high-capacity batteries, organic solar cells, extreme-strength composites, and other fields.

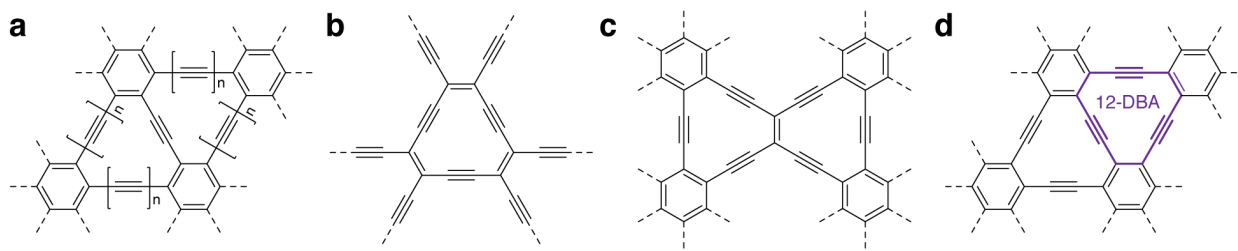

**Figure 1. Graphyne family allotropes of carbon. a**, Graphynes-*n* (graphdiyne for *n* = 2). **b**, 12,12,12-Graphyne. **c**, 6,6,12-Graphyne. **d**, Graphyne (or γ-graphyne).

In contrast, advances in the synthesis of nonbenzenoid carbon allotropes have been limited. Several sp$^2$-based nanographenes with non-hexagonal rings,[5, 6] as well as an extended nonbenzenoid biphenylene network,[7] were described. Materials containing sp$^1$ hybridized linear arrangements remain even more elusive. Linear polyyne chains -(C≡C)$_n$- are unstable even for moderate chain lengths unless they are stabilized with bulky end-capping groups[8] or confined inside carbon nanotubes.[9] Graphynes, a family of hybrid lattices combining sp$^1$ and sp$^2$ carbons, were first theoretically proposed by Baughman, Eckhardt, and Kertesz in 1987.[10] These allotropes can be formally viewed as graphenes that are expanded through insertion of acetylenic groups. Graphynes are commonly divided into two types: graphynes-*n* in which aromatic rings are separated by *n* acetylenic bonds (Fig. 1a), and *x,y,z*-graphynes featuring at least some non-aromatic sp$^2$ carbons (Fig. 1b and 1c). Theoretical studies of some graphyne lattices suggest unique electronic and chemical properties,[11] including an intrinsic band gap and electrochemical capacities that far exceed graphite.[12] To date, few-layer graphyne-2 or graphdiyne (Fig. 1a, $n = 2$) is the only graphyne family allotrope that has been synthesized (typically at sub-milligram scale) and thoroughly characterized.[13]

γ-Graphyne (Fig. 1d), the basic graphyne-$n$ homolog ($n = 1$), is especially intriguing. Single-layer γ-graphyne is predicted to be a semiconductor with a moderate band gap,[14] ultrafast charge carrier mobility comparable to that of graphene,[15] high thermal conductivity,[16] and exceptional strength.[17] Because of these properties, γ-graphyne could form the basis for the next generation carbon-based devices. γ-Graphyne oligomers containing up to four dehydrobenzo[12]annulene (12-DBA) repeat units have been prepared through multistep organic synthesis,[18,19] but synthesis of the extended crystalline lattice remains challenging.

Pyrolytic and vapor-deposition methodologies commonly used for the controlled synthesis of benzenoid carbons are unsuitable for $sp^1$-based structures, as acetylenes readily convert to graphene or amorphous carbon at elevated temperatures. While graphdiyne[13] and graphdiyne-graphene heterostructures[20] have been synthesized *via* templated solution-phase 2D polymerizations, a similar approach has not been previously attempted for γ-graphyne. Reported graphdiyne syntheses rely on polymerization of highly energetic hexaethynylbenzene through Glaser-Hay $sp^1$-$sp^1$ coupling, which can be conveniently localized at metal surfaces.[21] A comparable synthesis of γ-graphyne would require either coupling between $sp^1$ and $sp^2$ carbons, or *de novo* formation of three acetylenic bonds per each 12-DBA repeat unit. The most common and general methodology for $sp^1$-$sp^2$ C-C coupling is the Sonogashira reaction.[22] The mechanism of this reaction is thought to involve a homogeneous $Pd^0$ catalytic cycle, like all other Pd-catalyzed cross-couplings. Therefore, attempting to confine this chemistry to the surface of a template would be challenging. Moreover, previously reported γ-graphyne oligomers are distorted from planarity, due to steric hindrance introduced by the terminal functionalities.[23] This inevitable distortion, as well as the typically poor solubility of oligomers, limits stepwise extension of the lattice beyond 3-4 12-DBA units.[18]

**Results and Discussion:** Here, we demonstrate that under appropriately adjusted Sonogashira coupling conditions, an $A_3B_3$-type monomer, 1,3,5-tribromo-2,4,6-triethynylbenzene (TBTEB, Fig. 2a), can be polymerized into extended γ-graphyne. The main idea that guided our thinking is that an effective route to graphynes and similar rigid 2D polymers could proceed through reactions that create multiple connections in a single step or through a series of kinetically coupled fast steps. Such mechanism would bypass the kinetic dead-end of partially connected intermediates, as each monomer unit will "click" into place. Furthermore, this polymerization would be self-correcting. Defects in the growing lattice would be the most reactive sites due to local distortions and strain, which could be relieved upon multi-site reaction with the monomer. Thus, our two primary aims were to establish reaction conditions favoring exhaustive coupling of the multifunctional TBTEB, and to find a way to template the formation of the desired 2D lattice instead of disordered hyperbranched structures.

Several types of Suzuki-Miyaura, Kumada, and Negishi cross-couplings favor exhaustive substitution in multifunctional substrates. This mode of reactivity has previously been exploited for the syntheses of polyfunctional arenes[24] and low-defect hyperbranched polyphenylenes,[25] as well as for pseudo-living chain polymerizations.[26] In all these cases it is assumed that an exceptionally reactive Pd species is formed after the initial catalytic cycle.[24] The subsequent coupling steps are catalyzed by this Pd species, proceed inside the solvent cage, and are diffusion-controlled. We reasoned that we could extend this reaction mode to Sonogashira-type chemistry. Furthermore, while the nature of the Cu-mediated catalytic cycle in Sonogashira coupling remains largely unexplored, it is broadly understood to involve Cu acetylides.[22] The latter can assume either three-dimensional or low-dimensional polymeric forms[27] or possibly associate with a metal surface. We hypothesized that a Cu surface could template the γ-graphyne lattice, like it presumably does in the reported syntheses of graphdiyne.[13]

We synthesized TBTEB using a known method[28] and screened its reactivity under a range of conditions (Fig. 2a and Table S1, SI). In the control experiment in the absence of catalysts, TBTEB decomposed in refluxing pyridine with a half-life of ~24 hours, yielding an amorphous carbonaceous material (Fig. S19, SI). X-ray photoelectron spectroscopy (XPS) survey indicated moderate loss of Br through spontaneous hydrodebromination (Fig. S12 and S14d, SI). A similar featureless carbon was produced in the control experiment with just a Pd pre-catalyst and no source of Cu (Fig. S18, SI).

Experiments performed in the presence of both Pd and Cu produced outcomes dependent on the state of the metals. Pd(II) pre-catalysts, as well as PEPPSI-IPr, which we selected for its propensity for multi-site coupling,[24] yielded amorphous carbons broadly comparable to the control products. However, for stoichiometric loading of Pd(PPh$_3$)$_4$ in the presence of Cu foil, we obtained a black lustrous material (Fig. S15b, SI). Transmission electron microscopy (TEM) and scanning electron microscopy (SEM) images of this material revealed flakes composed of stacks of flat sheets (Fig. S17 and S21, SI). Selected area electron diffraction (SAED) experiments produced dotted ring patterns (Fig. S17e-f, SI), and no Moiré fringes were observed in bright-field TEM, indicating sub-micron crystalline domains with random orientation.

While the possibility of the layered flakes being a phase of γ-graphyne was intriguing, the data were insufficient to make a structural assignment. XPS survey indicated that the product was primarily carbonaceous, but contaminated by Pd, P, and C from the catalyst (Fig. S10, SI). Since a multilayered material was obtained, we reasoned that the coupling reaction cannot be confined to the surface of the foil and decided to investigate sources of Cu other than the metallic surface. To our surprise and delight, reactions employing soluble CuI also yielded layered flakes (Fig. 2b and S15a, S16, SI). The crystallinity of this material was significantly improved over the product produced using Cu foil, with Moiré fringes observable in bright-field TEM (Fig. S16a, SI). Some of the flakes had well-defined hexagonal shapes (Fig. 2b and S16d, SI). Micron-size hexagonal prisms with terraces could be observed in SEM (Fig. 2c and S20d, SI). A similar hexagonal morphology was previously reported for graphdiyne produced by interfacial synthesis.[29]

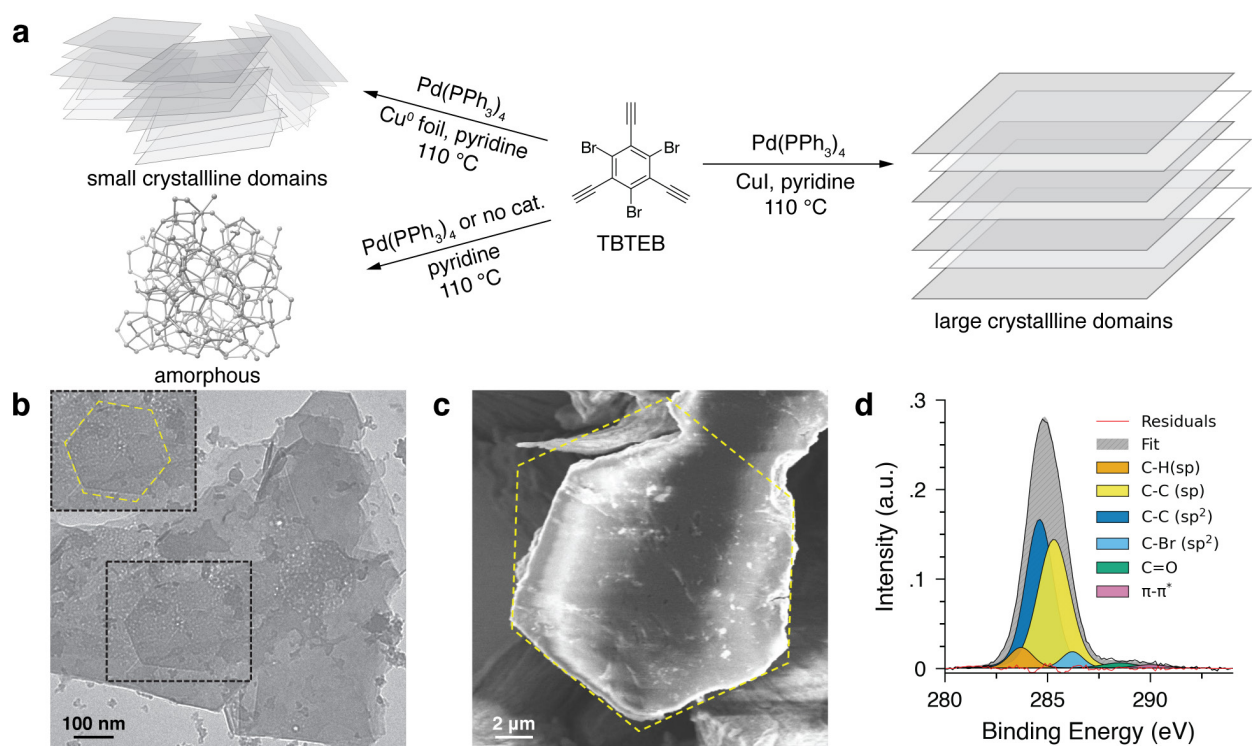

**Figure 2. Two-dimensional polymerizations of TBTEB. a**, Overview of selected reaction conditions. **b-c**, Representative bright-field TEM and SEM of the carbon flakes obtained from Pd(PPh$_3$)$_4$/CuI homogeneous reaction. Inset in **b** emphasizes the hexagonal shape of the layer. **d**, High resolution XPS for the C1s region of the sample in **b**.

The higher crystallinity of the product obtained through the optimized homogeneous Pd(PPh$_3$)$_4$/CuI protocol allowed for more efficient removal of contaminants. Survey XPS of this product indicated a level of contamination with Pd and P that was below the detection limit of the technique (Fig. S9, SI). We acquired high-resolution XPS data for the C1s region of this material (Fig. 2d), as well as for three controls: the product of the Cu foil synthesis, the Pd-only reaction, and thermal reaction products (Fig. S13, SI). The C1s peak can be deconvolved into five sub-peaks, corresponding to C-H (terminal alkyne $sp^1$),[30] C≡C (internal alkyne $sp^1$), C=C (aromatic $sp^2$), aromatic C-Br, and C=O carbons.[31] The contribution of C=O is negligible for all samples, indicating little to no oxidation under the reducing/anaerobic reaction conditions. Without the contribution of the $sp^1$ subpeak, none of the fits converge, which strongly supports the presence of acetylenic bonds in all products. XPS indicates a 1:1 ratio of $sp^1$ to $sp^2$ carbons in the crystalline material synthesized by the homogeneous $Pd(PPh_3)_4$/CuI protocol, which is consistent with γ-graphyne. This ratio is much higher in the control samples (Fig. S13b-d, SI), due to extensive side reactions and contamination with aromatic impurities. The π-π* "shake up" peak at 290 eV is commonly observed in XPS of graphitic carbons and graphene, as well as small aromatic molecules.[31] Notably, this peak does not appear in the XPS of graphdiyne.[20] The "shake up" feature was negligible for the product of the homogeneous Cu protocol (Fig. 2d), strongly suggesting that this material is not graphitic. The peak was prominent for the control products that were also contaminated with P (Fig. S10, S11, and S13c-d, SI), indicating that it may be related to adsorbed $PPh_3$.

The initial structural identification of the crystalline carbon material was made *via* synchrotron X-ray powder diffraction (PXRD) using a 0.728 Å wavelength (Fig. 3a). We observed a peak at 7.0° 2θ, which matches the predicted[11] 5.96 Å spacing between (10$\bar{1}$0) planes of γ-graphyne (Fig. 3a, left inset and Fig. 3d). The intense peak at 12.0° 2θ could be indexed as the (0003) plane, corresponding to an interlayer distance of 3.48 Å (Fig. 3a, right inset). To index the other observed peaks, we explored the possible crystal structures of γ-graphyne. While there is a single stable crystallographic configuration for two graphene sheets, a variety of arrangements are possible for bilayer γ-graphyne. Some of these bilayer stackings have been previously identified.[32] To systematically survey the possible structures, we used density functional theory (DFT) to analyze the potential energy surface for a γ-graphyne bilayer. The computed surface (Fig. 3e) is a function of the horizontal offset of the upper γ-graphyne layer relative to the lower layer with a fixed interlayer distance of 3.35 Å. The chosen interlayer distance was based on first-pass optimization of the bilayer geometry and is underestimated due to the difficulty of computationally modeling van der Waals interactions.[32] The calculations identified two types of local energy minima: one where the upper layer aromatic rings overlay the 12-DBA rings of the lower layer (Fig. 3e, binding site B), and the second one, where there is a fixed lateral distance between the centers of the upper and lower layer aromatic rings (Fig. 3e, binding site A). The absolute energy minimum of the former arrangement gives rise to a single crystal structure with AB mode of stacking corresponding to the P6$_3$mc space group. However, at least six bilayer structures are possible for binding at site A with energy minima that are nearly identical within the error of the calculations, giving rise to a multitude of AB, ABC, or more complex arrangements for multiple sheets. To further explore this complex energy map, which would be expensive for DFT calculations, we ran fully atomistic reactive molecular dynamics (MD) simulations for 3-6 layers of γ-graphyne. In all cases, the simulations converged to stacks of sheets bound exclusively at A sites (Fig. S34 and S35, SI). However, for multiple sheets stacked through A sites, the energy barrier for transitioning between different A site configurations is extremely small, since this energy difference depends upon the van der Waals interactions between non-adjacent graphyne sheets. Modeling indicated that several of the less-ordered stacking modes identified by the MD and DFT calculations produce PXRD patterns that closely match our experimental diffraction patterns. Thus, we could index the peak at 14.5° 2θ as the ($\bar{2}$112) plane.

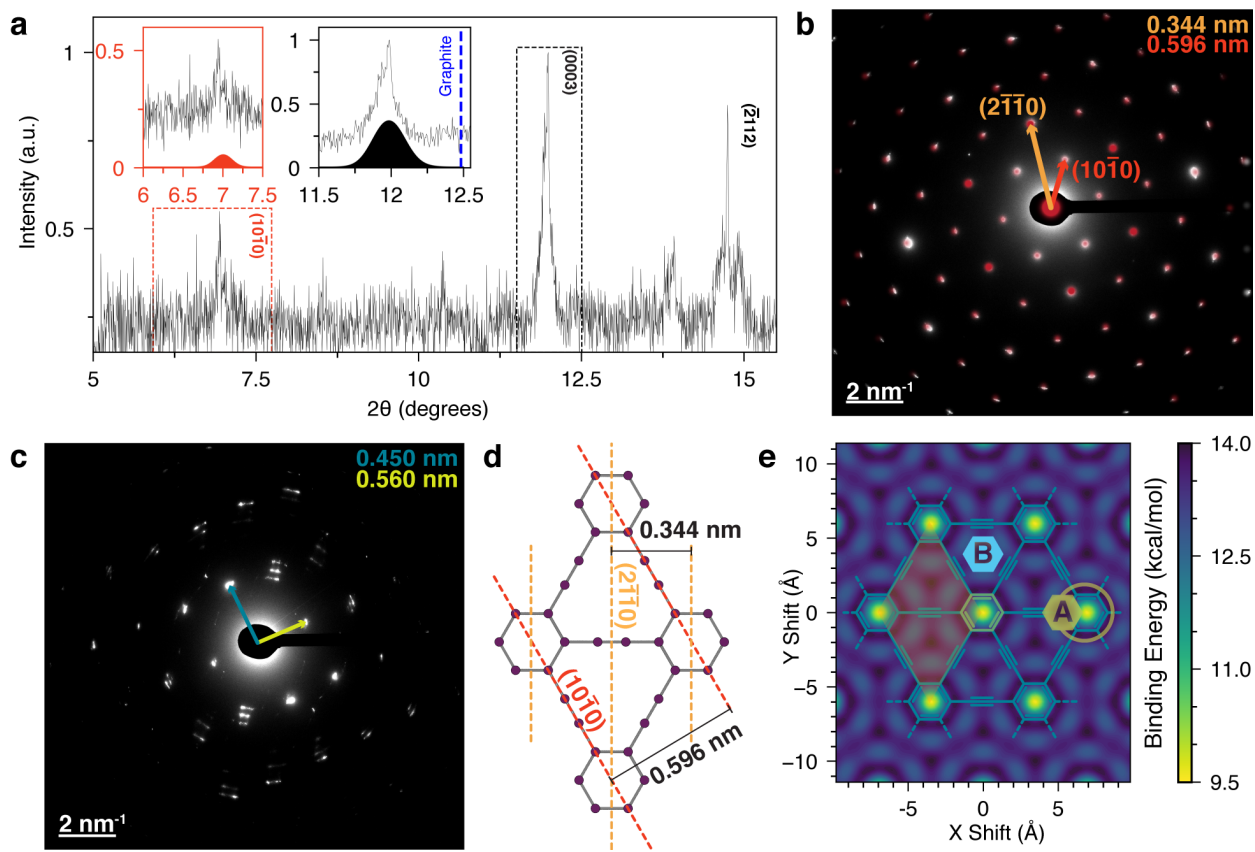

**Figure 3. X-ray and electron diffraction patterns of γ-graphyne. a**, Synchrotron PXRD pattern (0.728 Å radiation) of γ-graphyne produced by the Pd(PPh$_3$)$_4$/CuI protocol. Left inset (red): peak at 7.0° 2θ superimposed with the modeled (10$\bar{1}$0) peak for γ-graphyne with P3$_1$12 stacking. Right inset (black): peak at 12.0° 2θ superimposed with the (0003) peak for the same model. The dashed blue line is the (0002) peak center for graphite. **b**, Representative SAED (white) of the same material overlaid with the simulated (red) *c* orientation diffraction pattern for a structure with no systematic absences. **c**, SAED of the sample region in **b**, rotated by 45°. **d**, DFT-generated model of a γ-graphyne sheet overlaid with crystal planes of interest. **e**, DFT-generated potential energy surface for the stacking of two γ-graphyne sheets. The binding energy calculated for a single unit cell (highlighted).

We further explored the structure and symmetry of the crystals using electron diffraction. The spot SAED patterns of γ-graphyne produced with the Pd(PPh$_3$)$_4$/CuI protocol were exceptionally well-defined, consistent for different regions of the sample, and independent of the size or presence of a selected area aperture. The near-perfect uniformity of the patterns indicates that the material consists of crystalline domains that are sufficiently large to span the entire illuminated region of our typical imaging frame of 2×2 μm (Fig. 2b and Fig. S16, SI), indicating crystalline domain sizes of at least 1-3 μm.[33] The diffraction patterns observed from the flat areas of the sample had perfect hexagonal symmetry (Fig. 3b). Using bond distances calculated by DFT and the interlayer distance obtained from PXRD, we built models for several plausible stacking modes of γ-graphyne. These models were used to simulate electron diffraction in the *c*, *b*, and intermediate crystal orientations that lie ~45° to the (0001) pole (Fig. S23, SI). The simulated *c* orientation patterns exclusively involve spacings in the basal plane (Fig. S22c and S22d, SI). The first- and second-order reflections in the experimental diffraction pattern correspond to d-spacings of 5.96 Å and 3.44 Å, which perfectly match the theoretically calculated spacings[11] for the $(10\bar{1}0)$ and $(11\bar{2}0)$ plane sets of γ-graphyne (Fig. 3b and 3d). Both of these distances are defined in part by the length of the acetylenic bond. Some of the more symmetric space groups, such as Cmcm and R3m, are expected to produce diffraction patterns with systematic absences. Since there were no such systematic absences in the observed diffraction pattern, these space groups could be conclusively eliminated.

Additional SAED patterns were obtained for an alternate sample orientation. The initial position of the stage was chosen to yield the most symmetric spot intensity distribution, which corresponds to a beam normal to the basal plane and coincident with the *a* axis. Then the sample was rotated around the *b* axis. As the sample rotation reached ~45°, diffraction patterns that involve the *z* spacings began to appear. The experimental diffractograms in this orientation provided groups of closely spaced spots (Fig. 3c), suggesting defects in the layer stacking. Such defects would not appear in the *c* orientation diffractograms, since the stacking mode only affects reflections involving *z* spacing (Fig. S24, SI). The symmetry of the patterns agrees with our simulations for this intermediate orientation (Fig. S23e, 23f, 23h, SI). As no systematic absences were observed, we can exclude some of the more symmetric space groups, most notably the P6₃mc space group. The experimental diffraction patterns were most consistent with either one of the lower symmetry stacking modes, such as P3₁12 (Fig. S22e, SI) or an aperiodic superlattice. It is important to note that despite their multi-spot character, the observed diffractograms are not indicative of turbostratic stacking, which would produce ring patterns. The shape-factor effect alters the geometry of the diffracted beam,[34] which introduces error into the determination of spot centers. Although this prohibits precise measurement of interplanar spacings, the simulations indicate that the interlayer spacing estimated from SAED agrees with our PXRD data (Fig. S23, SI).

We further probed the structure of γ-graphyne flakes by using lateral force microscopy (LFM). LFM is a technique closely related to contact-mode atomic force microscopy (AFM). In LFM, the scanning tip is rastered across the surface of the sample while maintaining contact, and the torsional moment due to stick-slip friction is measured. LFM can achieve resolution approaching that of scanning tunneling microscopy (STM),[35] which is often the technique of choice for direct imaging of atomic arrangements. However, STM cannot be usefully applied to our sample due to the presence of absorbates that roughen its surface. In contrast, lattice-resolution LFM imaging can often be performed even on rough substrates under ambient conditions.[36] We prepared a sample for imaging by ultrasonicating the γ-graphyne flakes in water and casting them on a freshly cleaved surface of highly oriented pyrolytic graphite (HOPG). A lateral deflection map for a flat region of one of the γ-graphyne flakes (Fig. 4a-b) was then recorded. In the same imaging session, with the same tip, we then obtained a lateral deflection map of the underlying HOPG as a reference (Fig. 4d). Fast Fourier transform (FFT) of LFM images of both the γ-graphyne flake and HOPG showed hexagonal arrangements of high intensity spots, indicating hexagonal symmetry for both lattices (Fig. S27b-c, SI). The ~2.4 Å periodicity observed for the HOPG lattice (Fig. 4e, Fig. S26 and S27c, SI) agrees with the expected value.[35] The lattice of the γ-graphyne flake is visibly less dense (Fig. 4c). Its ~3.4 Å (one half of the $a$-axis spacing) periodicity agrees with the theoretically predicted distance between the $(11\bar{2}0)$ planes of γ-graphyne, which is also observed as a second-order reflection in electron diffraction (Fig. 3b, 3d, and 4f).

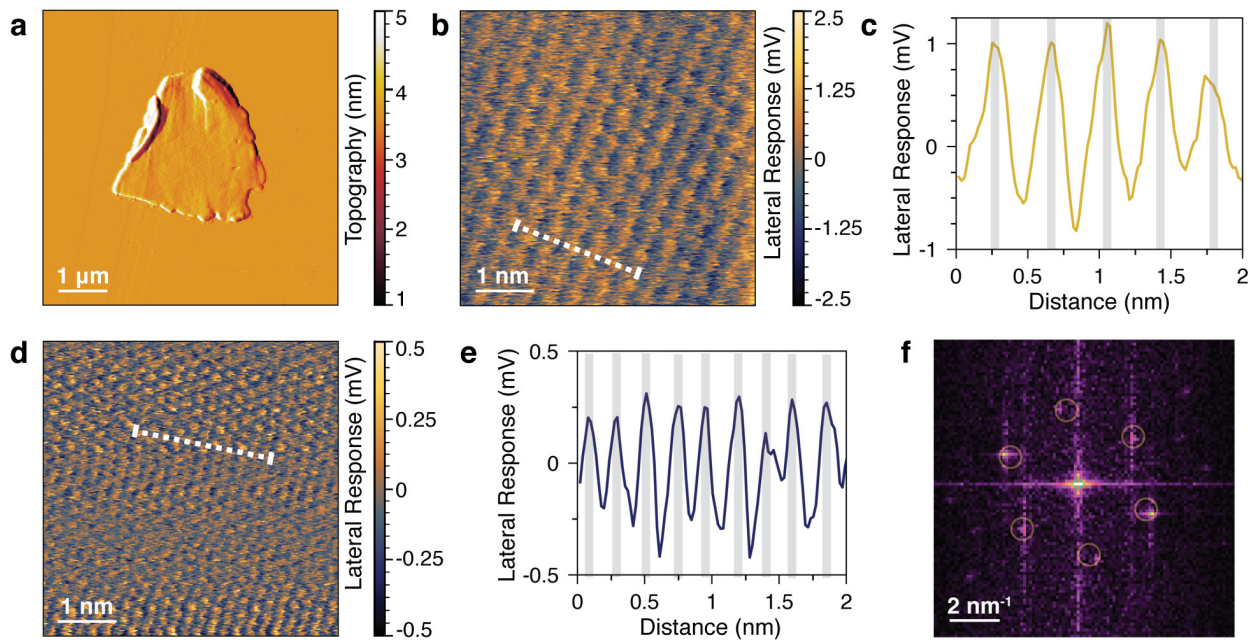

**Figure 4. Scanning probe microscopy of a γ-graphyne flake. a**, AFM topography map of a γ-graphyne flake. **b** and **d**, Lateral deflection maps of a region of the γ-graphyne flake and the HOPG substrate control in **a**. **c** and **e**, Plots of lateral response along the linear traces from **b** and **d**. Grey bars mark the confidence interval for lattice constants. **f**, Fast Fourier Transform (FFT) of the map from **b**. Circles highlight the periodicity for the $(2\bar{1}\bar{1}0)$ planes of γ-graphyne from electron diffraction (Fig. 3b).

The observed Raman spectra are consistent with expectations for γ-graphyne.[37, 38] The distinctive Y ($A_{1g}$) band corresponding to the C≡C stretch of internal triple bonds appears at 2197 cm$^{-1}$ (Fig. 5a). The G band, which corresponds to the $E_{2g}$ modes of the aromatic rings, is centered at 1565 cm$^{-1}$, exhibiting the predicted softening[37] relative to the G bands of common graphitic materials (~1580 cm$^{-1}$) due to additional resonant configurations from π-electron delocalization along the acetylenic linkages.[39] A broad D band, corresponding to $A_{1g}$ breathing of aromatic rings, is observed at ~1350 cm$^{-1}$ and is expected to be sensitive to domain size, lattice defects, and the excitation wavelength. A broad survey scan (100 – 3000 cm$^{-1}$) showed no other Raman features for the 405 nm excitation wavelength, most notably no C-H stretches (2800-3000 cm$^{-1}$) or the Y′ band at ~1900 cm$^{-1}$ characteristic of diacetylenes.[29] In the spectra of the TBTEB monomer (Fig. S32, SI), a band at 2114 cm$^{-1}$, corresponding to the stretch of the terminal C≡C-H triple bonds, is observed but not seen in the spectra of γ-graphyne, most likely due to the low ratio of internal to terminal triple bond sites.

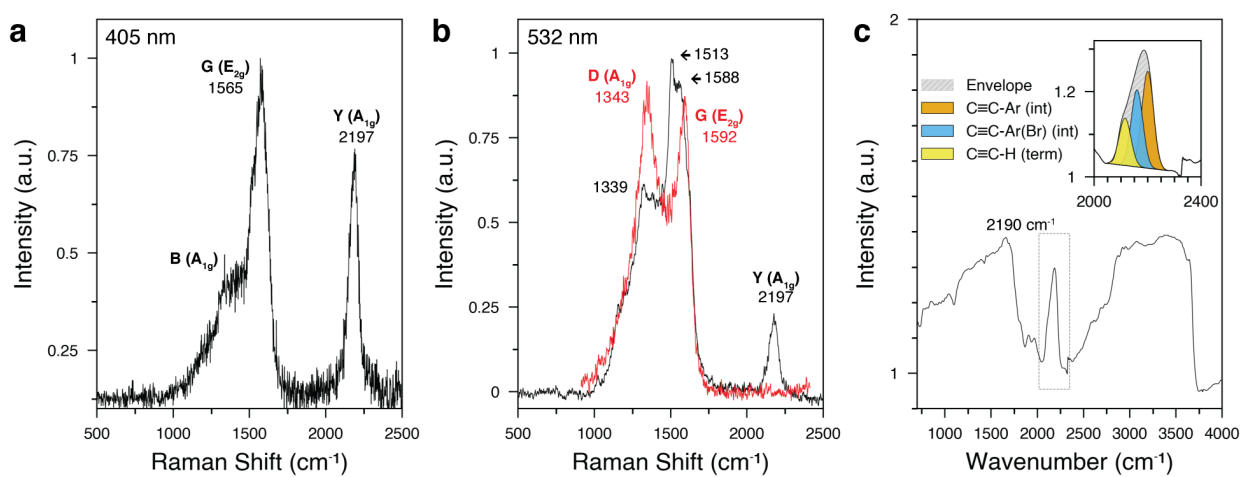

**Figure 5. Vibrational spectroscopy of γ-graphyne. a**, Raman spectrum of γ-graphyne measured using an excitation wavelength of 405 nm (4.3 mW). **b**, Raman spectra of γ-graphyne measured using an excitation wavelength of 535 nm. Black trace: <0.3 mW for less than 10 s. Red trace: <0.3 mW, after high-power irradiation at 7.5 mW for 30 s. **c**, Mid-IR spectrum of polycrystalline γ-graphyne. Inset: deconvolution of the alkyne absorption peak at 2190 cm$^{-1}$.

It is important to note that our attempts to obtain Raman spectra of γ-graphyne using 532 nm excitation, for above 0.75 mW incident power using a 0.55 NA 50× objective, resulted in rapid and irreversible transformation of the material. This transformation was marked by the bleaching of the Y band, the increase in the intensity of the D band, and the shift of the G band to ~1590 cm$^{-1}$ (Fig. 5b). The resulting Raman signature is similar to that of disordered graphitic carbons.[40] The transformation is not due to direct oxidation, since experiments conducted in air and under ~1×10$^{-3}$ mbar vacuum resulted in similar observations. γ-Graphyne retains its characteristic Raman signature under prolonged high power 405 nm illumination (>4 mW), while rapidly transforming under 532 nm laser light (Fig. S28c, SI), which suggests a photochemical process. Under stable low power 532 nm illumination (<0.3 mW), the degree of transformation does not linearly correlate with the exposure dose (Fig S28a-b, SI), suggesting thresholding photochemical behavior. A plausible pathway of the transformation involves Masamune-Bergman cycloaromatization, which has been demonstrated for a single 12-DBA subunit on a copper surface.[41] The observed splitting of the G band under 532 nm excitation, but not 405 nm excitation (black trace, Fig. 5b) is not yet well understood, and could be due to partial transformation from the 532 nm light even when measured at a relatively low excitation intensity.

The micro-FTIR spectra of polycrystalline flakes of γ-graphyne featured wide-band absorption in the fingerprint region, as well as between 2800-3700 cm$^{-1}$ (Fig. 5c). Absorption in these regions is associated with vibrations of the aromatic rings of γ-graphyne, as well as contributions from the C-H and O-H groups on the periphery of the sheets or belonging to adventitious small-molecule adsorbates. The distorted line shape of these bands, as well as the change in line shape observed for flakes of different thicknesses suggests significant contribution from resonant Mie scattering.[42,43] Thus, we could not identify specific vibration frequencies or their true intensities for low- and high-frequency mid-IR regions. However, a distinctive peak corresponding to acetylenic bonds centered at ~2190 cm$^{-1}$ was observed in the mid-IR silent region (1700-2500 cm$^{-1}$). Since the C≡C stretch is IR-inactive for symmetric alkynes, an ideal infinite monolayer of γ-graphyne would not possess this band. However, symmetry breaking due to stacking of graphyne sheets, as well as finite and defective sheets, could activate this absorption, like for the appearance of the D band in the Raman spectra of milled graphite.[44] The IR absorption reveals a prominent low-frequency shoulder (Fig. 5c, inset), due to the contribution of the terminal alkyne species at sheet edges (C≡C-H stretch, ~2115 cm$^{-1}$).

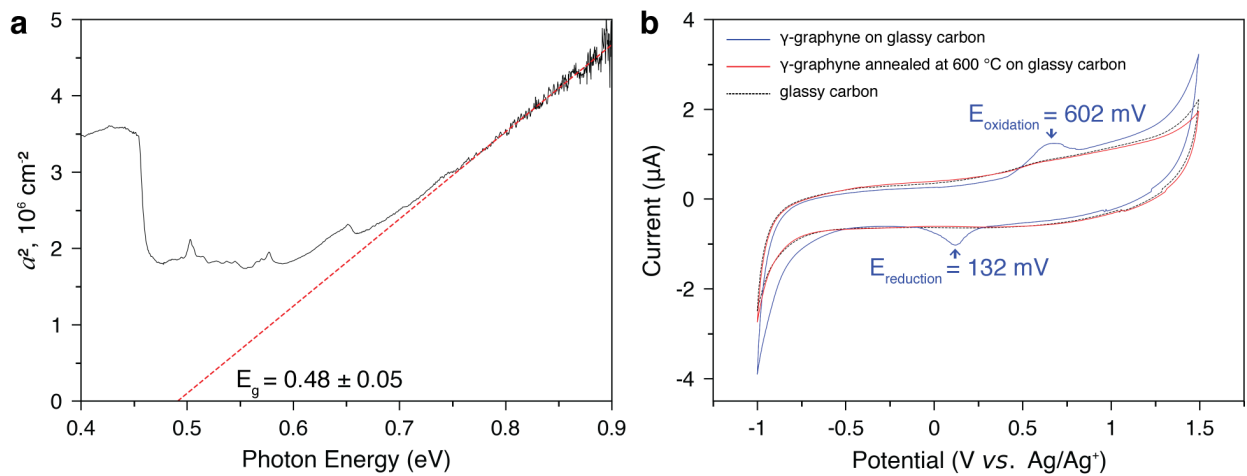

**Figure 6. Electronic properties of γ-graphyne. a**, Determination of the optical bandgap from the near-IR spectrum of polycrystalline γ-graphyne. **b**, Cyclic voltammetry of γ-graphyne powder on glassy carbon. These are three-electrode measurements in 0.1 M $n$-Bu$_4$N·PF$_6$/acetonitrile supporting electrolyte, with Pt counter electrode and an Ag/AgNO$_3$ (0.1 M) reference electrode at a scan rate of 50 mV/s.

The near-IR absorbance spectrum of γ-graphyne shows a strong onset of the fundamental electronic edge at high frequencies (5000-7500 cm$^{-1}$), which is manifested as a monotonic decrease of absorbance with respect to frequency, and as an optical phonon fingerprint in the infrared shoulder (Fig. 6a). The dependence of absorbance on photon energy within the electronic edge can be used to estimate the electronic band gap, $E_g$, and determine the type of semiconductor. The absorption coefficient α is defined as α = (1/$x$)×ln(1/$T$), where $x$ is the sample thickness and $T$ is transmittance, the ratio of transmitted to incident light intensity. Within the semi-classical theory of the optical absorption of crystalline direct band gap semiconductors, α($E$) is expected to be 0 for $E < E_g$, and proportional to $(E - E_g)^{1/2}$ for $E \geq E_g$.[45, 46] Thus, $E_g$ can be estimated by plotting α$^2$ versus $E$, and extrapolating the linear region of the curve to the energy axis (Fig. 6a). This α$^2$ versus $E$ plot gives an optical band gap of $E_g = 0.48 \pm 0.05$ eV for γ-graphyne, where the uncertainty indicates the standard deviation for over 20 polycrystalline particles.

To investigate the redox properties of γ-graphyne, a working electrode was prepared by ultrasonically dispersing 1 mg of γ-graphyne in 1 mL of a 1:1 v/v mixture of ethanol and water, and then casting this suspension onto a glassy carbon electrode with subsequent air drying. Three-electrode cyclic voltammetry (CV) measurements were then conducted in acetonitrile with $n$-Bu$_4$N·PF$_6$ supporting electrolyte for potentials between -1.0 and +1.5 V $vs.$ Ag/Ag$^+$ (0.1 M). An oxidation peak at 602 mV and a reduction peak at 132 mV were observed (Fig. 6b, blue trace), corresponding to an electrochemical bandgap of 0.47 eV.[47] This value agrees with the bandgap determined by the optical absorption measurements.

To evaluate thermal stability, γ-graphyne particles were supported over holes in thin mica films by single-layer carbon nanotube sheets and then heated to gradually increasing set points under vacuum (1.3×10$^{-3}$ mbar). After one hour at the set temperature, the sample assembly was cooled down and transferred to the FTIR microscope for analysis. The same sample region was analyzed for consistency. Heating up to 240 °C did not produce any major changes in FTIR spectra beyond a slight decrease in the intensity of the 2800-3700 cm$^{-1}$ band, likely due to the removal of adventitious adsorbates (Fig. S30a, SI). Loss of the alkyne band at 2100-2200 cm$^{-1}$ was observed at just over 240 °C, suggesting an onset of a structural transformation. The material lost all the alkyne absorbance after the one-hour thermal cycling processes reached 350 °C (Fig. S30b, SI). The absorbance shoulder corresponding to terminal alkynes disappeared before the internal alkyne components of the band. The loss of alkyne absorption was accompanied by the disappearance of the Mie scattering bands in the fingerprint region and the region between 2800-3700 cm$^{-1}$, suggesting significant rearrangement of the microcrystalline structure. The material annealed to 600 °C lost the characteristic reduction and oxidation peaks in CV, indicating a major chemical change (Fig. 6b, red trace). Despite the dramatic transformation of the IR signature, the optical images of the thermally treated samples revealed no textural or geometric changes on heating up to 450 °C (Fig. S30c, SI). Due to its sensitivity to the different alkyne bonds present in the sample, the IR absorbance was found to be a useful marker for monitoring the structural changes of γ-graphyne.

Our data indicates that the material we synthesized is multilayer γ-graphyne. Contrary to expectations, we found that no external template is required for synthesizing highly crystalline γ-graphyne. There is no experimental evidence even in our reactions performed with Cu foil that any polymerization is happening on the surface. The lower crystallinity of the Cu foil products is likely due to the reduced concentration of catalytic Cu species in solution. This results in slower Sonogashira coupling and a higher extent of side reactions compared to the homogeneous Pd(PPh$_3$)$_4$/CuI protocol.

We tried to understand why TBTEB preferentially polymerizes into multilayer γ-graphyne flakes, as opposed to amorphous branched structures. As the polymerization appears to be self-templating, we assumed the existence of an attractive supramolecular interaction between TBTEB and the lattice of γ-graphyne. Self-assembly through solvophobically driven π-stacking has been documented for several phenylene ethynylene oligomers and macrocycles structurally related to graphynes.[48] To explore the potential supramolecular interactions in our system, we computed by DFT the structure and potential energy surface for a single TBTEB molecule bound to a γ-graphyne monolayer. Our calculations predict that TBTEB would associate with the surface of graphyne at two types of binding sites (around the aromatic rings and over 12-DBA rings) with a binding energy in excess of 20 kcal/mol (Fig. S33, SI). The monomer species outcompetes toluene, whose binding energy we estimate as ~11 kcal/mol. If every TBTEB species were to react while bound over the underlying layer of γ-graphyne, one of the many local energy minimum stackings or a mixture of the stackings could result.

**Conclusions**: To our knowledge, our synthesis of γ-graphyne is the first example of an ordered covalent lattice formed spontaneously under purely kinetic control. Typical covalent organic frameworks and metal-organic frameworks are held together through bonds that are reversible. This reversibility is considered critical for continuous error correction during the reaction/crystallization process.[49] Conventional thinking predicts that irreversible polymerization of an $A_3B_3$-type monomer, not employing a strict geometric constraint on reactivity, must yield only disordered branched structures. However, since we routinely observed micron-scale γ-graphyne crystallites, 2D polymerization assisted by crystallization must be presently kinetically favored over random 3D growth. Furthermore, the initial nucleation of flat graphyne sheets appears to be a highly probable event. The high fidelity of the resulting lattices indicates that the system is capable of correcting errors despite the irreversibility of Sonogashira coupling. At a minimum, the reaction must proceed comparably well at both lattice edges and internal defect sites. Since "patching" a single internal defect requires forming six new chemical bonds, it is highly likely that these bond-making steps are kinetically coupled. The apparent capability for error correction, as well as the strong dependence of the product structure on the nature of the Pd pre-catalyst, strongly corroborate our original hypothesis of a multi-site coupling mechanism.

Similar cross-coupling methodology could conceivably be applied to the synthesis of other graphyne-family allotropes, as well as to theoretically proposed heteroatom-doped derivatives.[50] Performing the reaction at interfaces may provide access to extended few-layer or monolayer sheets rather than microcrystalline powders. These extended sheets could form the basis of the first γ-graphyne-based devices, especially since we observe a small, direct band gap. Further exploration of the chemical and physical properties of γ-graphyne is under way in our laboratories.

ASSOCIATED CONTENT

Materials and methods, experimental procedures, NMR and FT-IR spectra, crystallographic information for possible stacking modes of γ-graphyne sheets, supplemental figures and discussion referred to in the text, and computational data (.PDF)

Models of possible stacking modes of γ-graphyne sheets (.CIF)

AUTHOR INFORMATION

**Corresponding Author**

*Correspondence to: Prof. V.O. Rodionov. Email: vor2@case.edu

**Author Contributions**

‡These authors contributed equally.

ACKNOWLEDGMENT

We are grateful to Prof. Jessica Bickel for discussions of scanning probe microscopy data. We thank the US Department of Energy (R01AB123456) and the National Science Foundation (GFRP Award 1451075 to WBM) for funding. RHB acknowledges support from Robert A. Welch Foundation (grant AT-0029). DSG and AFF acknowledge support from São Paulo Research Foundation (FAPESP, awards #2013/08293-7 and #2020/02044-9), National Council for Scientific and Technological Development (CNPq), and the John David Rogers Computing Center (CCJDR) at the Institute of Physics "Gleb Wataghin", University of Campinas. We thank Oxford Instruments Asylum Research for providing access to Cypher VRS AFM instrument. This work made use of the High Performance Computing Resource in the Core Facility for Advanced Research Computing at Case Western Reserve University. Raman spectroscopy was performed at the Stanford Nano Shared Facilities (SNSF), supported by the National Science Foundation under award ECCS-2026822. The use of the Stanford Synchrotron Radiation Lightsource, SLAC National Accelerator Laboratory, was supported by the U.S. Department of Energy, Office of Science, Office of Basic Energy Sciences, under Contract DE-AC02-76SF00515.

Supplementary Materials for

# Scalable Synthesis and Characterization of Multilayer γ-Graphyne, New Carbon Crystals with a Small Direct Band Gap

Victor G. Desyatkin,[1]‡ William B. Martin,[1]‡ Ali E. Aliev,[2] Nathaniel E. Chapman,[1] Alexandre F. Fonseca,[3] Douglas S. Galvão,[3] Ericka Roy Miller,[4] Kevin H. Stone,[5] Zhong Wang,[2] Dante Zakhidov,[6] F. Ted Limpoco,[7] Sarah R. Almahdali,[1] Shane M. Parker,[4] Ray H. Baughman,[2] and Valentin O. Rodionov[1]*

‡These authors contributed equally.

[1]Department of Macromolecular Science and Engineering, Case Western Reserve University; 2100 Adelbert Road, Cleveland, OH 44106, USA.

[2]Alan G. MacDiarmid NanoTech Institute, University of Texas at Dallas; 800 West Campbell Road, Richardson, TX 75080, USA.

[3]Applied Physics Department, Institute of Physics "Gleb Wataghin", University of Campinas; Campinas, SP, 13083-970, Brazil.

[4]Department of Chemistry, Case Western Reserve University; 10900 Euclid Ave, Cleveland, OH 44106, USA.

[5]Stanford Synchrotron Radiation Lightsource, SLAC National Accelerator Laboratory; 2575 Sand Hill Road, Menlo Park, CA 94025, USA.

[6]Department of Materials Science and Engineering, Stanford University; 496 Lomita Mall, Stanford, CA 94305, USA.

[7]Oxford Instruments Asylum Research; 6310 Hollister Ave, Santa Barbara, CA 93117, USA.

*Correspondence to: Prof. Valentin O. Rodionov. Email: vor2@case.edu

# Materials and Methods

Materials

All reagents and solvents were acquired from commercial suppliers (Acros Organics, Sigma-Adrich, TCI Chemicals, Fisher Scientific, Oakwood Chemical and VWR International) and used without further purification, unless otherwise noted. Tetrahydrofuran (THF) was distilled over Na/benzophenone. Triethylamine (TEA) was distilled over $CaH_2$. Anhydrous pyridine (Py) was purchased from Acros in AcroSeal packaging and used without further purification. Copper foil, 1.0 mm thick Puratronic 99.999% (metals basis) with 25x50 mm lateral dimensions, was purchased from Alfa Aesar and cut into 5x10 mm pieces. These pieces of were then sequentially sonicated for 20 minutes in 3M HCl, water, ethanol, and acetone, dried under high vacuum at ambient temperature, and immediately used.

Synthetic Methods

Reactions were monitored by thin-layer chromatography (TLC) carried out on 0.25 mm MilliporeSigma aluminum-backed silica gel plates (60F-254). Plates were visualized using 254 nm UV light and basic potassium permanganate stain (1.5 g $KMnO_4$, 0.5 g NaOH, and 10 g $K_2CO_3$ in 150 ml water; terminal alkynes stain yellow). Flash chromatography was performed on Luknova SuperSepTM (230-400 mesh) silica gel. Reactions requiring anhydrous or air-free conditions were performed under positive pressure of $N_2$ or Ar using standard Schlenk line techniques.

Nuclear Magnetic Resonance (NMR) Spectrometry

NMR spectra were recorded on a Bruker Avance III HD 500 spectrometer operating at 500.24 ($^1$H), 125.79 ($^{13}$C), or 99.37 ($^{29}$Si) MHz and equipped with Bruker Ascend 500 MHz US Narrow Bore Magnet and Broadband Prodigy TCI CryoProbe. NMR spectra were referenced to TMS ($^1$H, $^{13}$C, $^{29}$Si) or residual solvent peaks. Chemical shifts (δ) are reported in parts per million (ppm).

Gas Chromatography – Mass Spectrometry

GC-MS analyses were performed on an Agilent 5977B GC/MSD instrument equipped with an Agilent 7890B automatic liquid sampler. Before injection of the sample, the 10 μL syringe was cleaned with acetone and ethyl acetate (3x10 µL each). 1 µL of the sample was then automatically injected into the instrument. The method used a 3-minute solvent delay. The oven was initially set at 60 °C and held at this temperature for 2.25 minutes before increasing the temperature to 225 °C at 35 °C/min rate. Data analysis was performed using Agilent MassHunter Qualitative Analysis Navigator.

Infrared Spectroscopy

Routine small molecule FTIR spectra were collected on an Agilent Cary 630 FTIR instrument equipped with single-reflection germanium or diamond attenuated total reflectance (ATR) modules. The instrument was calibrated before sampling against a newly cleaned (acetone) and dried crystal surface. Solid samples were placed directly on the crystal and secured with a needle press. 32 scans from 4000 to 550 cm$^{-1}$ were recorded. A background was collected for each sample (512 scans).

We attempted to obtain FTIR spectra of γ-graphyne using the routine spectrometer described above. When using the diamond ATR module, we observed primarily artifacts related to the diamond substrate itself (Fig. S29a), suggesting that the refractive index (RI) of γ-graphyne in mid-IR is higher than that of diamond ($n = 2.4$). Some IR bands in the fingerprint region (600-1800 cm$^{-1}$) and at ~2100 cm$^{-1}$ could be observed using a germanium ATR crystal ($n = 4.0$) (Fig. S29b). However, this spectrum featured low intensity and unfavorable signal-to-noise ratio because of the low penetration depth for ATR FTIR on germanium. Likewise, we could not obtain satisfactory data using the KBr pellet technique due to the low absorbance.

Therefore, for all subsequent experiments we utilized a PerkinElmer Spotlight 200i FTIR Microscopy System equipped with a Spectrum Two spectrometer capable of both transmittance and reflectance measurements in the mid-IR to near-IR range (600-7800 cm$^{-1}$).

The micro-FTIR spectra were collected at ambient conditions in reflection mode using an 8 cm$^{-1}$ resolution, 50x50 µm$^2$ aperture, and 100 scans. The typical polycrystalline particles chosen were on average ~300×300 µm and 20 µm thick. Spectra was deconvolved using CasaXPS software.[1,2] The peaks were modeled using a Gaussian/Lorentzian (SGL(10)) sum formula SGL(p) where the mixing between the two peak types are determined by m = p/100. SGL(100) is pure Lorentzian while SGL(0) is entirely Gaussian. A linear background was used.

Raman Spectroscopy

Raman spectra were obtained using a Horiba Labram HR Evolution confocal Raman microscope with a 405 and 532 nm excitation laser source in reflection geometry. An Edwards T-Station 85 turbo pump connected to a Janis ST-500 stage was used for vacuum control. Samples were measured in vacuum at a pressure reading of $1\text{-}2\times10^{-5}$ mbar to control for possible oxidation. The vacuum monitoring was done with a pressure gauge situated next to the turbo pump. It is most likely that the pressure in the Janis ST-500 stage was closer to $1\times10^{-3}$ mbar. A 50x 0.55NA Mitutoyo long working-distance objective (13 mm) was used to collect Raman spectra from diffraction-limited spot sizes of ~620 nm (532 nm laser) and ~470 nm (405 nm laser). A low laser power of less than ~0.3 mW for the 532 nm laser was used for ideal measuring conditions to reduce beam damage and inhibit decomposition. Powers up to 3.0 mW were used to explore laser light induced transformations. For the 405 nm laser, laser powers between 0.43 and 4.3 mW were used. Multiple spots were measured on each sample and no differences were found in the Raman spectra.

Melting Points

Melting points were determined with a Mettler Toledo MP50 Melting Point System.

Preparation of Exfoliated Samples

Analyte dispersions in water (1 mg/mL) were prepared by ultrasonication using a Branson SFX550 Sonifier. A 1/8" double-step microtip (Branson p/n 101-063-212) was used. Samples were processed for 15 minutes at 50% amplitude.

Scanning Electron Microscopy (SEM)

The SEM images of Fig. S20a-b and S21 were acquired on an FEI Apreo 2 SEM operating at 5kV and an FEI Inspect F-50 operating at 30kV, respectively. For these images, dried material was added to carbon tape and mounted on SEM sample stands and sputtered with a thin layer of gold.

The SEM images of γ-graphyne in Fig. 2 (Main Text) and Fig. S20c-d were acquired by a high resolution Zeiss Supra-40 system, using an in-lens detector and a 10 kV accelerating voltage.

The samples for these images were not coated with gold. The small γ-graphyne particles were grounded to the gold-covered silicon substrate using silver paste.

Transmission Electron Microscopy (TEM) and Selected Area Electron Diffraction (SAED)

Exfoliated samples were analyzed by TEM. Prior to sample preparation, 200-Cu C-B grids were plasma-treated for 30 seconds using an Emitech K100x glow discharger. 3 µL of sample dispersion was added to the grid and allowed to absorb for 5 minutes before the excess solvent was wicked. The grid was then transferred to a single-tilt sample holder and imaged on an FEI Tecnai 20 TEM operating at 200 kV in low-dose mode. Images were recorded on a Tvips F416. Data was collected using SerialEM software. Tilting was performed with the equipped alpha-rotation goniometer.

SAED patterns were recorded on an FEI Tecnai 20 TEM using a 40 µm selected area aperture and in the absence of the selected area aperture. The obtained patterns were calibrated against the (111) planes of evaporated aluminum (plane spacing 0.2338 nm) on a 3 mm grid. The calibration sample was purchased from Electron Microscopy Sciences (EMS p/n 80044).

X-ray Photoelectron Spectroscopy (XPS)

Samples were spread onto double-sided copper tape for XPS analysis. Surveys and high-resolution spectra were acquired on a PHI VersaProbe II Scanning XPS Microprobe using a monochromatic Al X-ray at pressures of $10^{-10}$ to $10^{-7}$ Torr. The data was smoothed by using the Savitzky-Golay method, with a smoothing width of five, and analyzed using CasaXPS software.[1,2]

A Tougaard background[3] was applied to each peak before deconvolution. All peak fits used generalized Voigt-like peak shapes, as this function is most appropriate for fitting asymmetric XPS signals.[4] CasaXPS provides a generalized Voigt function described as Lorentzian Finite: LF(α, β, w, n, m), where the first three parameters (α, β, w) affect the Lorentzian line shape and its asymmetry and the final two (n, m) change the width of the Gaussian function and the number of times convolution with the Lorentzian component occurs.[5] Symmetrical peak parameters for the LF line shape were used: LF(1, 1, 255, 360, 6), values derived from default symmetric peak shape settings for CasaXPS. All sub-peak widths were constrained to full width at half maximum (FWHM) of 1.6 eV or less. The subpeaks are located at 283.7 eV (terminal alkyne $sp^1$),[6] 284.6 eV (aromatic $sp^2$), 285.3 eV (internal alkyne $sp^1$), 286.9 eV (aromatic C-Br), and 288.5 eV (carbonyl C=O).[6, 7] All peaks were allowed a ± 0.2 eV padding to the peak position.

Synchrotron Powder X-ray Diffraction

Powder diffraction data were collected at the Stanford Synchrotron Radiation Lightsource (SSRL, SLAC National Accelerator Laboratory) beamline 2-1. Samples were loaded into thin-walled glass capillaries of 0.5mm nominal diameter. The capillaries were spun during data collection to improve powder averaging. Measurements were made using 0.7281 Å wavelength X-rays. Diffracted X-rays were collected using a Pilatus 100K area detector mounted approximately 700 mm from the sample on the 2-theta arm of the beamline diffractometer. Each image covers an angular range of approximately 5 degrees. Images were collected 0.0625 degrees apart, providing substantial overlap between images, and subsequently merged and integrated into the final powder diffraction profile using custom software available at the beamline.

Scanning Probe Microscopy

Scanning probe microscopy was performed on an Oxford Instruments Asylum Research Cypher VRS AFM instrument equipped with an Olympus RC800PSA probe with a nominal spring constant of 0.39 N/m. Images were taken at 5 nm scan size, 16 Hz line rates, and 256 x 256 pixels. The analyte powder was suspended in HPLC grade water and cast on freshly cleaved highly oriented pyrolytic graphite (HOPG). The sample flakes were located optically using the AFM camera and the tip precisely positioned on the material using the scanner as inertial motors. Lateral force microscopy (LFM) was the imaging mode used to obtained lattice resolution images. In this mode, the tip is rastered orthogonal to the long axis of the cantilever. The torsional moment of the cantilever due to stick-slip friction was registered as the lateral signal in the photodetector, which is sensitive to the lattice corrugation of the surface.

# Computation and Modeling

Density Functional Theory (DFT) Calculations

All DFT calculations were performed using 2D periodic boundary conditions via the RIPER module of TURBOMOLE/7.5.[8-12] In all cases the PBE density functional[13] with D3 dispersion corrections[14] and Becke-Johnson damping[15] was used. All the calculated geometries were in broad agreement with prior computational studies of γ-graphyne.[16-18]

Potential energy surfaces of both a γ-graphyne bilayer and individual molecules/graphyne supercells were computed as a function of the horizontal offset of the upper γ-graphyne layer or molecule relative to the lower layer with a fixed interlayer distance of 3.35 Å. Binding energies for each structure were calculated as adsorption energies ($E_{binding} = E_{bilayer} - E_{separated}$). A 9x9 k-point grid was used for the γ-graphyne bilayer structures. The monomer/graphyne supercells were built using a 3x3 γ-graphyne monolayer to ensure 1 nm spacing between adjacent periodic images of monomers. Due to the resulting repetitiveness of these supercells, a coarser 3x3 k-point grid was used throughout the potential energy surface scan calculations. The def2-SVP basis set [19] was used throughout.

Local bilayer minima identified from the potential energy surfaces were refined by geometry optimization with the def2-TZVP basis set[20] and a 9x9 k-point grid. Similarly, selected binding site supercell structures were optimized with def2-SVP and a 9x9 k-point grid. Final binding energies for all structures were calculated with def2-TZVP as well, along with a finer 17x17 k-point grid.

Molecular Dynamics (MD) Simulations

The energy and dynamics of a few different stackings of γ-graphynes were investigated using classical MD simulations. LAMMPS package[21] was used with the ReaxFF reactive force field.[22,23] ReaxFF is a state-of-the-art potential previously applied to simulations of structural, mechanical, and thermal properties of carbon nanostructures,[24-28] including graphynes.[29-32]

The following γ-graphyne stackings were used as initial structures: AA, AB1 and AB2, and ABC (Fig. **S35**). AB1 and AB2 were previously proposed by Yun.[33] AB1 and ABC structures were investigated by Ducéré and Chauvin[34] as the most probable stackings of γ-graphyne. However, as we will see below, neither corresponds to the lowest energy stacking mode.

All the structures were prepared with 3 and/or 6 layers and were first energy-minimized and then either equilibrated at 300 K or quenched from 1000 to 1 K. The energy minimizations were performed with periodic boundary conditions (PBC) imposed along all directions in space, including the possibility to relax the size along the PBC directions. Combination of energy minimization with free evolution algorithms as suggested by Sihn[35] was used to ensure the lowest-energy structure is obtained. This protocol was recently used by Kanegae and Fonseca[36] to study elastic properties of graphynes. Thermal equilibration of the structures was performed with PBC by applying a Langevin thermostat to all atoms with damping factor of 1.0 fs and timestep of 0.025 fs, for a total period of 500 ps (or $20 \times 10^6$ timesteps). Quenching of the structures was simulated at the same conditions as thermal equilibration except that a Boltzmann distribution of velocities corresponding to the initial temperature of 1000 K was initially attributed to the system, then the temperature was allowed to decrease from 1000 to 1 K over a period of 500 ps.

SAED and PXRD Simulations

The lattice parameters and bond lengths were obtained from DFT calculations (*vide supra*) and a previously published computational study.[16] SAED and PXRD simulations were performed using the CrystalMaker software suite.[37] A model of a single γ-graphyne sheet was built in CrystalMaker using a hexagonal P6 lattice with parameters *a* and *c* set to 6.86 Å and 3.4 Å, respectively. The asymmetric unit comprised four atoms placed at 0.208, 0.412, 0.589, 0.795 along the hexagonal P6 *x* axis. The basic models corresponding to various sheet stacking modes were constructed using Vesta.[38] Models shown in Fig S22c and S22e-h were generated from DFT simulations and were passed through Avogadro/spglib[39] in an attempt to identify the space group associated with each model.

# Small Molecule Synthesis

1,3,5-tribromo-2,4,6-triiodobenzene

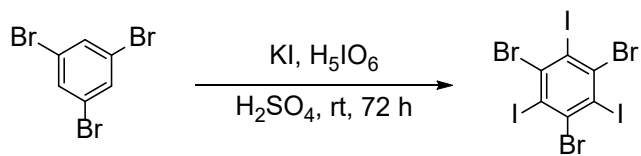

The synthesis procedure for 1,3,5-tribromo-2,4,6-triiodobenzene was adapted from the literature.[40] To concentrated $H_2SO_4$ (500 mL) at room temperature was added periodic acid (41.03 g, 180 mmol) in small portions over 15 min. After dissolution of the periodic acid, crushed KI (89.64 g, 540 mmol) was added in small portions at 0 °C over 1 h. To the resulting deep purple solution at 0 °C was added 1,3,5-tribromobenzene (18.89 g, 60.0 mmol) in small portions over 25 min. After the solution was stirred at room temperature for 72 h, the resulting thick mixture was poured onto ice. The resulting precipitate was filtered and washed with $H_2O$ (5 × 200 mL) and then MeOH (5 × 200 mL). The product was recrystallized twice from pyridine/EtOH 1:4 (1000 mL) to yield a solid. The solid was dried at under high vacuum for 1 day to give 1,3,5-tribromo-2,4,6-triiodobenzene 2 (28 g, 67%) as a pale-yellow solid. Mp > 300 °C (decomposition); FTIR (neat) $\nu_{max}$ = 1488, 1354, 1262, 1227, 1147, 1002, 858, 771, 739, 554, 508 cm$^{-1}$. $^{13}$C NMR (126 MHz, DMSO-$d6$) δ 138.61 (CBr), 108.23 (CI). EI-MS fragmentation: m/z 695.5, 693.5, 691.5, 689.5, 567.6, 566.6, 565.6, 564.6, 439.7, 437.6. UV/vis (CHCl$_3$, C = 6.874 × 10$^{-5}$ M): $\lambda_{max}$ (ε) = 227 (5000), 248 (27700), 283 (5600 M$^{-1}$cm$^{-1}$).

((2,4,6-tribromobenzene-1,3,5-triyl)tris(ethyne-2,1-diyl))tris(trimethylsilane)

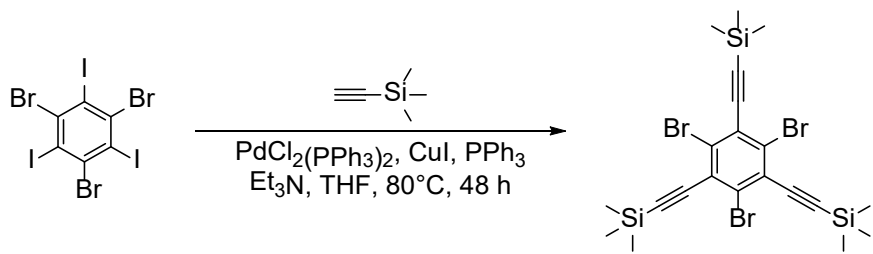

1,3,5-tribromo-2,4,6-triiodobenzene 2 (346 mg, 0.5 mmol), [PdCl$_2$(PPh$_3$)$_2$] (105 mg, 0.15 mmol, 30 mol%), CuI (19 mg, 0.1 mmol, 20 mol%), Et$_3$N (50 mL) and THF (40 mL) were added to a dry three-necked flask. Ethynyltrimethylsilane (736.7 mg, 1.07 mL, 7.5 mmol) and Ph$_3$P (52 mg, 0.2 mmol, 40 mol%) were added to the mixture. The mixture was stirred at 80 °C for 48 h under argon. After the removal of solvent on a rotary evaporator, DCM (100 mL) was added to the residue and filtered through Celite. The mixture was washed with water (20 mL) and NaCl(aq) (20 mL), dried over anhydrous Na$_2$SO$_4$, and the solvent was removed under reduced pressure. The residue was further purified by flash chromatography using n-hexane as the eluent to yield ((2,4,6-tribromobenzene-1,3,5-triyl)tris(ethyne-2,1-diyl))tris(trimethylsilane) 3 as a white solid (175 mg, 0.29 mmol, yield: 58%). R$_f$ (hexane) = 0.3. Mp = 110-111 °C; FTIR (neat) ν$_{max}$ = 2958, 2160, 1376, 1340, 1245, 1019, 834, 758, 708, 658, 633, 539 cm$^{-1}$. $^1$H NMR (500 MHz, CDCl$_3$): δ = 0.29 ppm [s, 27H, Si(CH$_3$)$_3$]. $^{13}$C NMR (126 MHz, CDCl$_3$): d 129.09 (CBr), 127.49 (C$_6$C≡C), 106.79 (C≡CSi), 101.83 (C$_6$C≡C), –0.23 [Si(CH$_3$)$_3$] ppm. $^{29}$Si NMR (99 MHz, CDCl$_3$) δ -15.87 ppm. EI-MS fragmentation: m/z 603.9, 602, 601.9, 590.9, 589.9. 588.9, 588.9, 587.9, 586.9, 584.9. UV/vis (CHCl$_3$, C = 5.436 × 10$^{-5}$ M): λ$_{max}$ (ε) = 260 (45700), 271 (44600), 289 (39600 M$^{-1}$cm$^{-1}$).

1,3,5-tribromo-2,4,6-triethynylbenzene, TBTEB

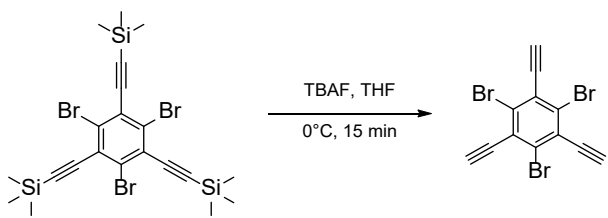

To a solution of ((2,4,6-tribromobenzene-1,3,5-triyl)tris(ethyne-2,1-diyl))tris(trimethylsilane) (151 mg, 0.25 mmol) in THF (15 mL) was added 0.55 mL TBAF (75% solution in water, 1.5 mmol) and stirred at 0 °C for 15 min. The solution was then diluted with ethyl acetate and washed with distilled water and dried with anhydrous Na$_2$SO$_4$. The solvent was removed on a rotary evaporator. The residue was further purified by flash chromatography using n-hexane as the eluent to give TBTEB as a white solid (84 mg, 0.216 mmol, yield: 87%). R$_f$ (hexane) = 0.2. FTIR (neat) ν$_{max}$ = 3275, 2922, 2112, 1519, 1368, 1336, 965, 736, 681, 634 cm$^{-1}$. $^1$H NMR (500 MHz, CDCl$_3$): δ = 5.16 (s, 3H, (C≡CH) ppm. $^{13}$C NMR (126 MHz, CDCl$_3$): δ 129.90 (CBr), 126.33 (C$_6$C≡CH), 91.87 (C≡CH), 80.97 (C$_6$C≡CH) ppm. EI-MS fragmentation: m/z 390.8, 389.9, 388.8, 387.8, 386.8, 385,8, 384,8, 383.8. UV/vis (CHCl$_3$, C = 4.136 × 10$^{-5}$ M): λ$_{max}$ (ε) = 212 (7100), 218 (8300), 260 (58100), 278 nm (33500 M$^{-1}$cm$^{-1}$).

# Synthesis of Carbon Materials

**Table S1.** Representative Polymerization Reactions of TBTEB.[a]

| Entry | [Pd] / mol% | [Cu] / mol% | Additive / mol% | Solvent / Base | Temp, °C | Time, h | Product |
|---|---|---|---|---|---|---|---|
| 1 | Pd(PPh$_3$)$_4$ / 100% | CuI / 8% | - | Pyridine | 110 | 72 | high crystallinity |
| 2 | Pd(PPh$_3$)$_4$ / 100% | Cu foil[1] | - | Pyridine | 110 | 48 | low crystallinity |
| 3 | Pd(PPh$_3$)$_4$ / 100% | - | - | Pyridine | 110 | 72 | amorphous |
| 4 | - | - | - | Pyridine | 110 | 72 | amorphous |
| 5 | PdCl$_2$(PPh$_3$)$_2$ / 30% | Cu/Si | Ph$_3$P / 40% | THF/ Et$_3$N (44:56) | 80 | 48 | amorphous |
| 6 | - | Cu/Si | - | Pyridine | 110 | 72 | amorphous |
| 7 | - | Cu wire | - | Pyridine | 110 | 72 | amorphous |
| 8 | Pd(PPh$_3$)$_4$ / 100% | Cu wire | - | Pyridine | 110 | 72 | low crystallinity |
| 9 | - | Cu wire[1] | - | Pyridine | 100 | 72 | amorphous |
| 10 | Pd(PPh$_3$)$_4$ / 40% | - | - | Pyridine | 100 | 48 | amorphous |
| 11 | Pd(PPh$_3$)$_4$ / 100% | Cu/Si | - | Pyridine | 110 | 72 | amorphous |
| 12 | - | Cu foil[1] | - | Pyridine | 110 | 72 | amorphous |
| 13 | PdCl$_2$(PPh$_3$)$_2$ / 60% | Cu/Si | Ph$_3$P / 80% | THF/ Et$_3$N (1:1) | 80 | 72 | amorphous |
| 14 | - | Cu foil | - | Pyridine | 70 | 72 | No Material |

[1]Copper foil was treated by sonicating in 3M HCl, water, ethanol, and acetone, sequentially, for 20 minutes, dried under vacuum at rt and used immediately.

[a] All the reactions were performed under positive pressure of Ar (Schlenk line with a Hg bubbler).

General Synthetic Procedure for Carbon Materials

In a typical procedure, TBTEB, Pd(PPh$_3$)$_4$, and Cu were charged to a Schlenk tube under argon atmosphere and solvent was added. The tube was sealed, and the contents degassed by three freeze-pump-thaw cycles. The reaction mixture was stirred under argon atmosphere and heated for 72 hours. The reaction mixture was concentrated on a rotary evaporator. The solid product was washed with methanol, ethanol, isopropanol, toluene, hexanes, ethyl acetate and acetone. The washing procedure involved dispersing the material in the corresponding solvent by gentle sonication, followed by centrifugation. Conditions for selected experiments from Table S1 are detailed below.

Table S1, Entry 1

TBTEB (116 mg, 0.3 mmol), Pd(PPh$_3$)$_4$ (347 mg, 0.3 mmol) and CuI (4.6 mg, 0.024 mmol) reacted in anhydrous pyridine (50 mL) using the general procedure.

Typical mass of the crude product after centrifugation and drying on low vacuum (1-2 Torr) over 10 hours is ~90% of the monomer mass (104 mg for the scale above). After extensive drying at high vacuum (10 mTorr) and/or heating to 100°C for 72 hours the mass decreased to ~60% of the original monomer mass (68 mg for the scale above). TLC indicated monomer conversion is quantitative.

Table S1, Entry 2

TBTEB (116 mg, 0.3 mmol), Pd(PPh$_3$)$_4$ (347 mg, 0.3 mmol) and several pieces of copper foil reacted in a mixture of anhydrous pyridine (50 mL) using the general procedure.

Table S1, Entry 3

TBTEB (116 mg, 0.3 mmol) and Pd(PPh$_3$)$_4$ (347 mg, 0.3 mmol) reacted in pyridine (50 mL) using the general procedure. No Cu was used.

Table S1, Entry 4

TBTEB (116 mg, 0.3 mmol) refluxed in pyridine (50 mL). Neither Pd(PPh$_3$)$_4$ nor copper was used.

# Spectroscopic and Imaging Data and Supplementary Discussion

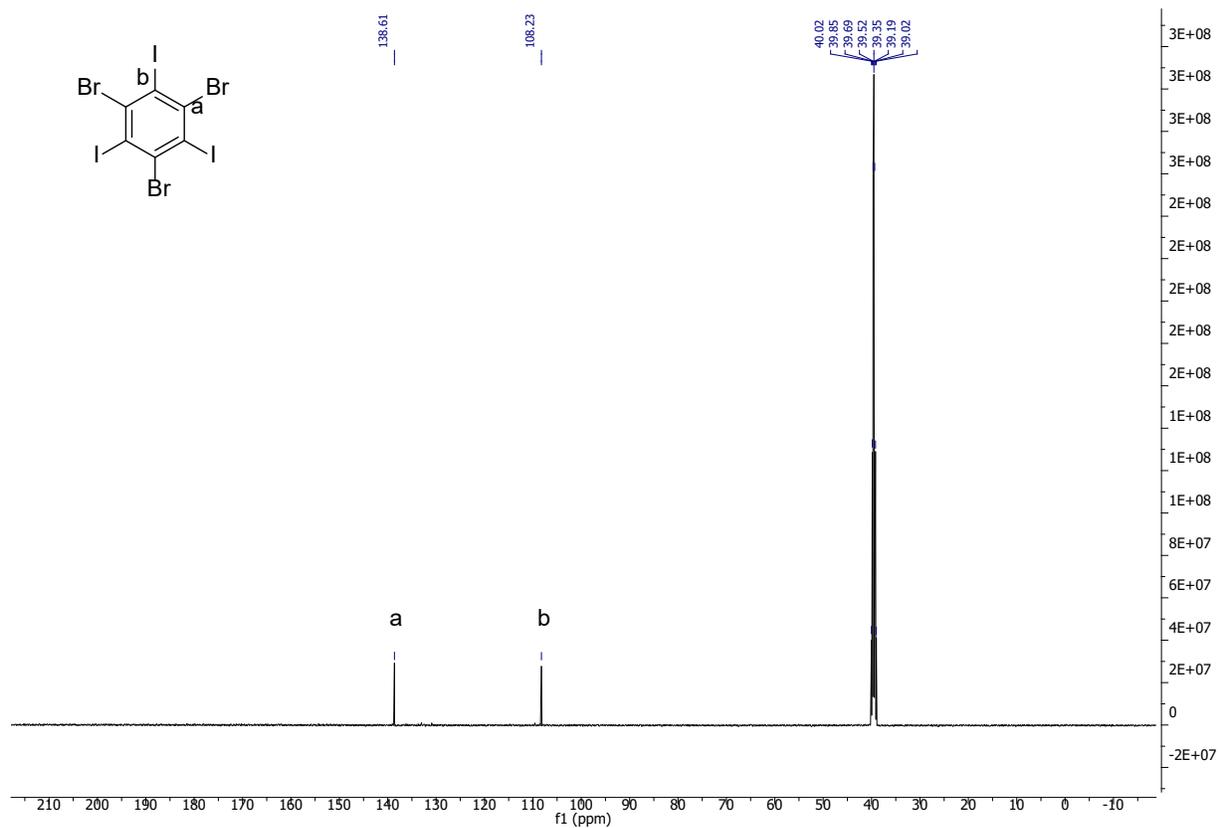

**Fig. S1.** $^{13}$C NMR spectrum of 1,3,5-tribromo-2,4,6-triiodobenzene in DMSO-d6.

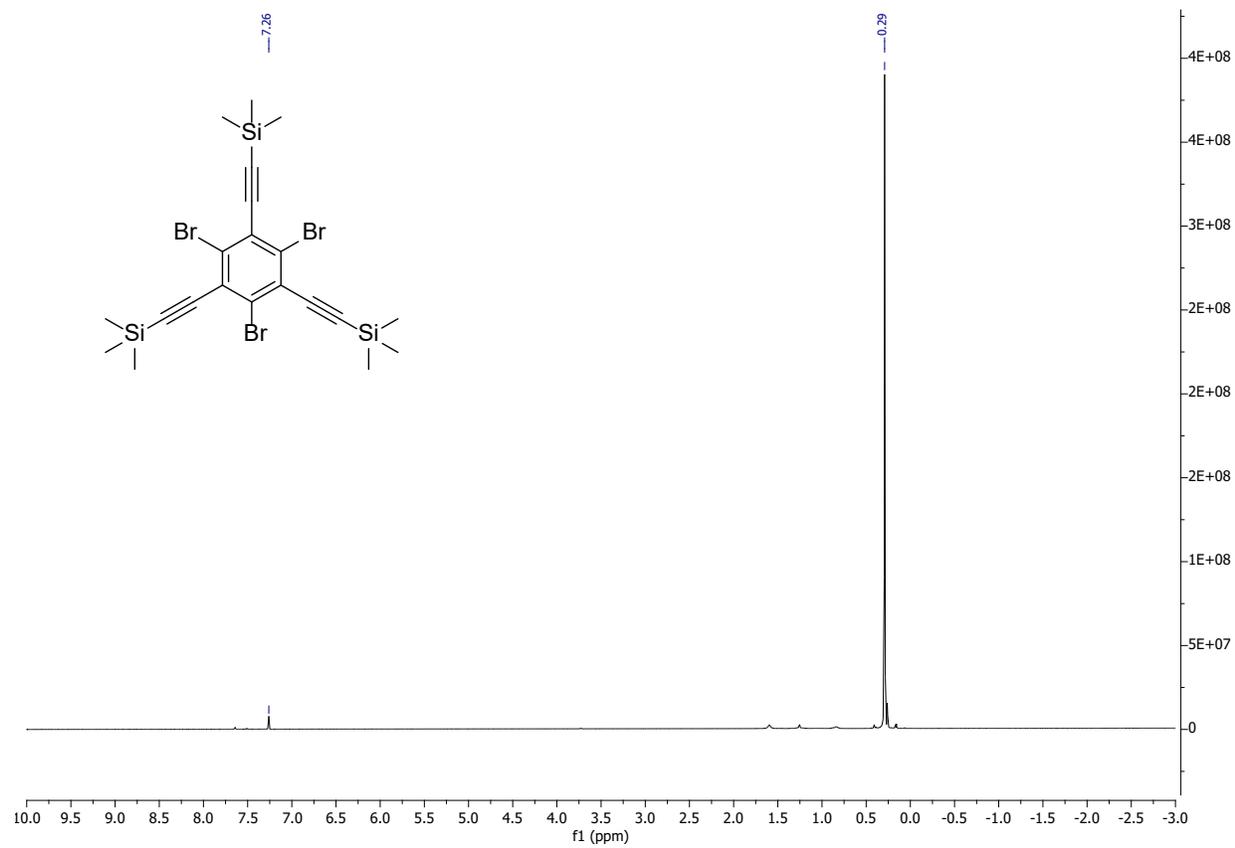

**Fig. S2.** [1]H NMR spectrum of ((2,4,6-tribromobenzene-1,3,5-triyl)tris(ethyne-2,1-diyl))tris-(trimethylsilane) in CDCl$_3$.

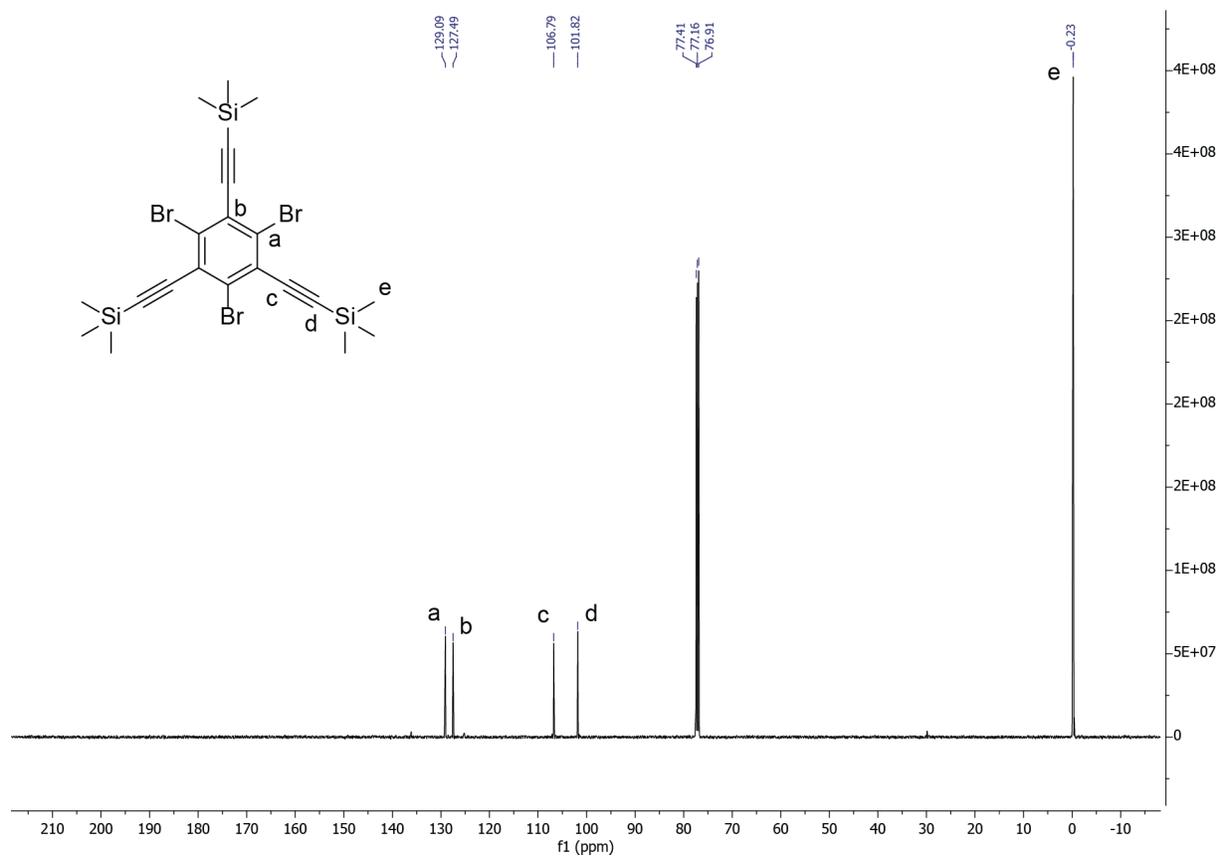

**Fig. S3.** $^{13}$C NMR spectrum of ((2,4,6-tribromobenzene-1,3,5-triyl)tris(ethyne-2,1-diyl))tris-(trimethylsilane) in CDCl$_3$.

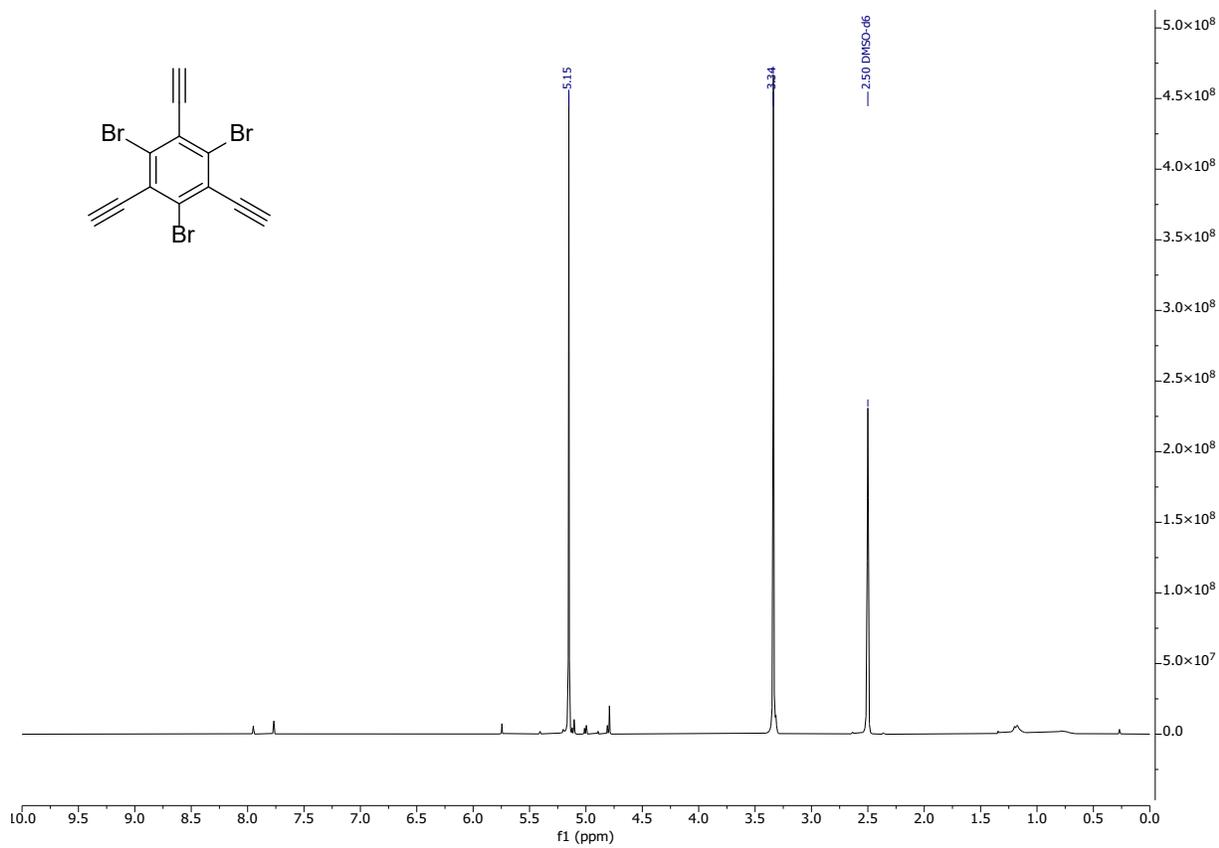

**Fig. S4.** ¹H NMR spectrum of 1,3,5-tribromo-2,4,6-triethynylbenzene, TBTEB in DMSO-d6.

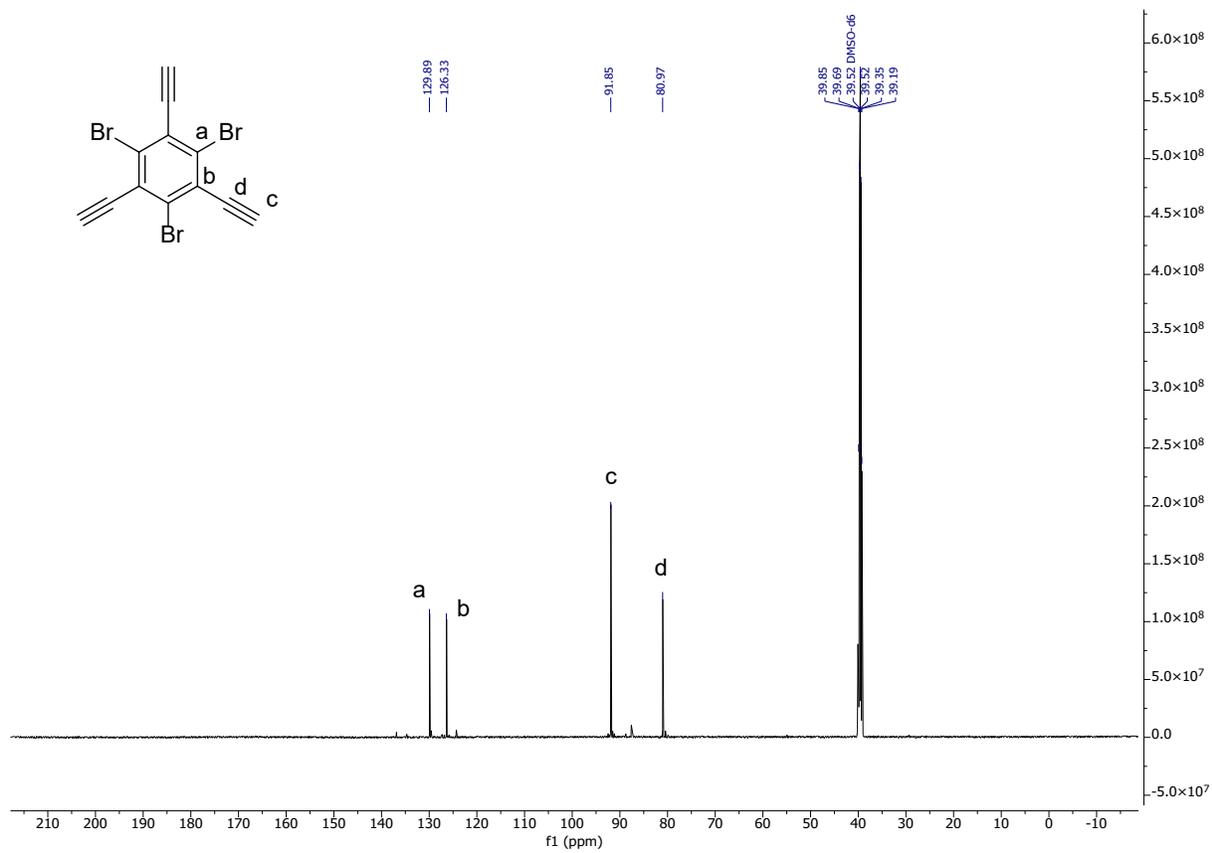

**Fig. S5.** $^{13}$C NMR spectrum of 1,3,5-tribromo-2,4,6-triethynylbenzene, TBTEB in DMSO-d6.

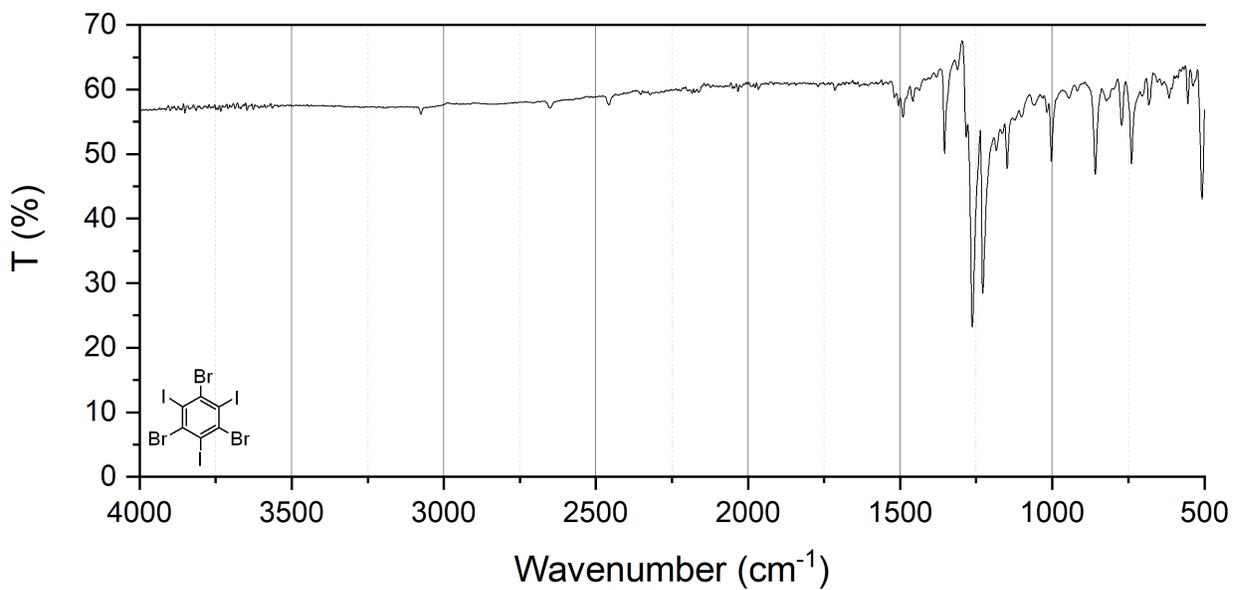

**Fig. S6.** FTIR of 1,3,5-tribromo-2,4,6-triiodobenzene (ATR-FTIR on diamond).

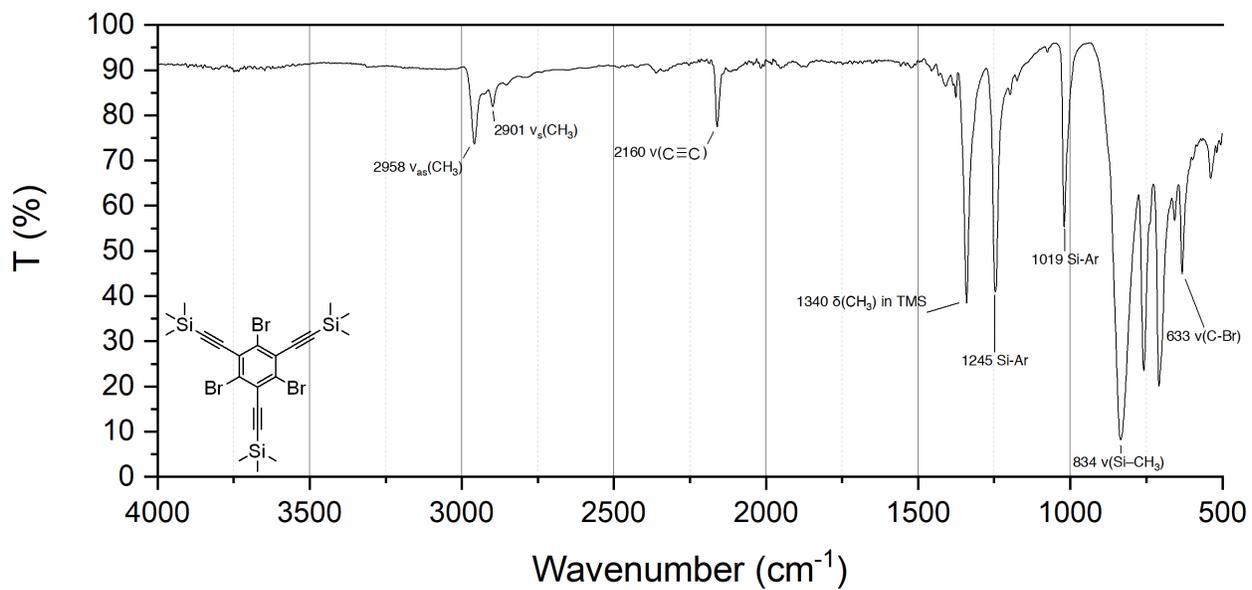

**Fig. S7.** FTIR of ((2,4,6-tribromobenzene-1,3,5-triyl)tris(ethyne-2,1-diyl))tris(trimethylsilane) (ATR-FTIR on germanium).

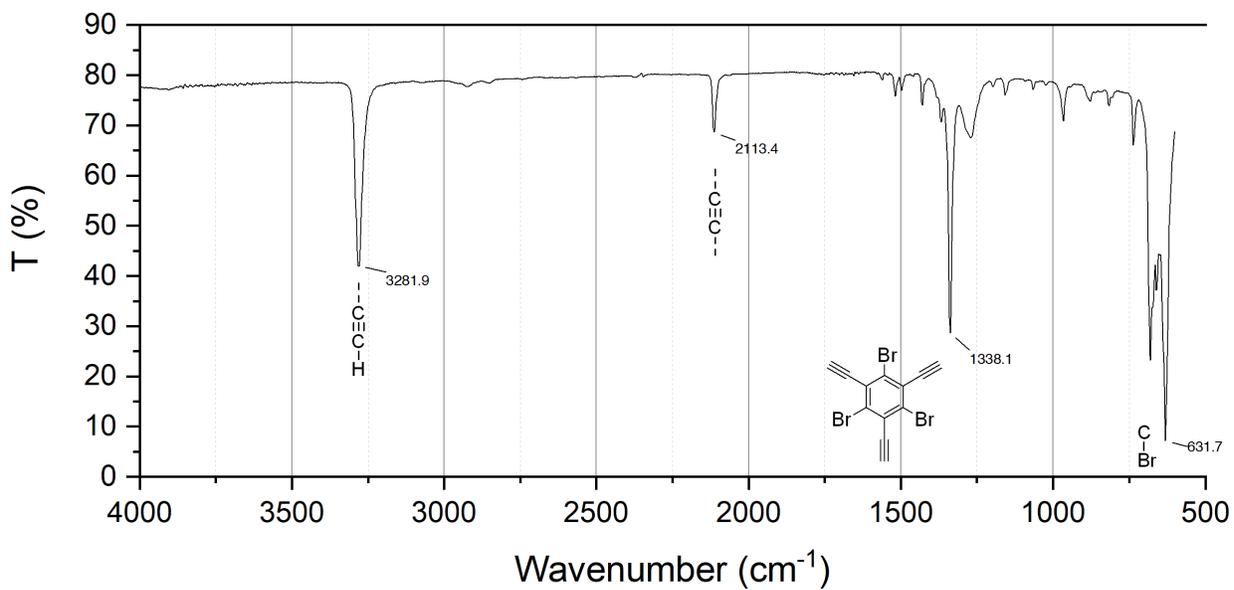

**Fig. S8.** FTIR of 1,3,5-tribromo-2,4,6-triethynylbenzene, TBTEB (ATR-FTIR on germanium).

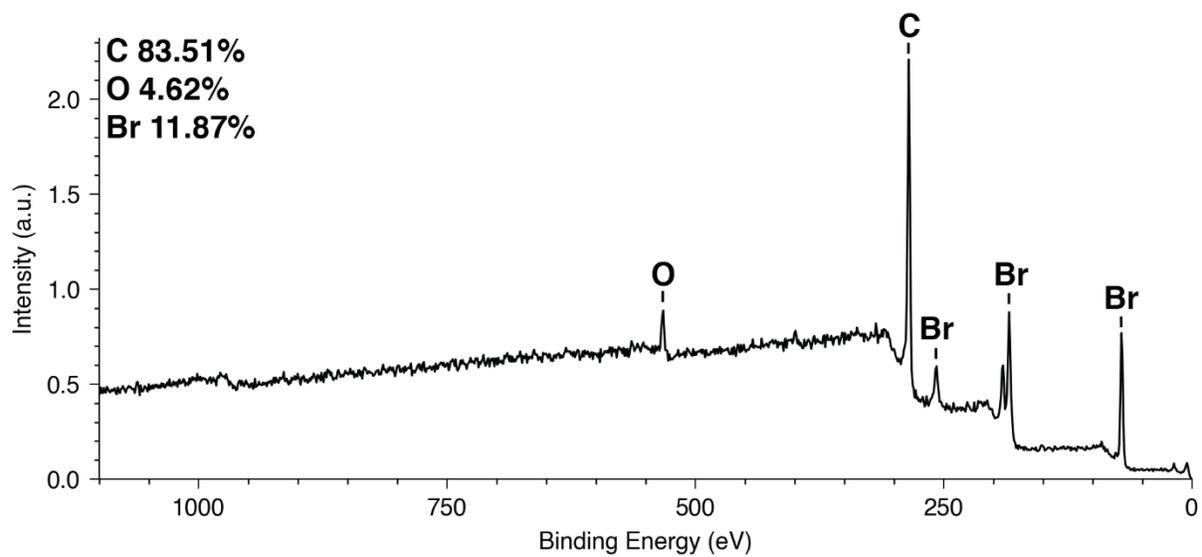

**Fig. S9.** XPS survey corresponding to Table S1, Entry 1 (TBTEB and Pd(PPh$_3$)$_4$/CuI in pyridine).

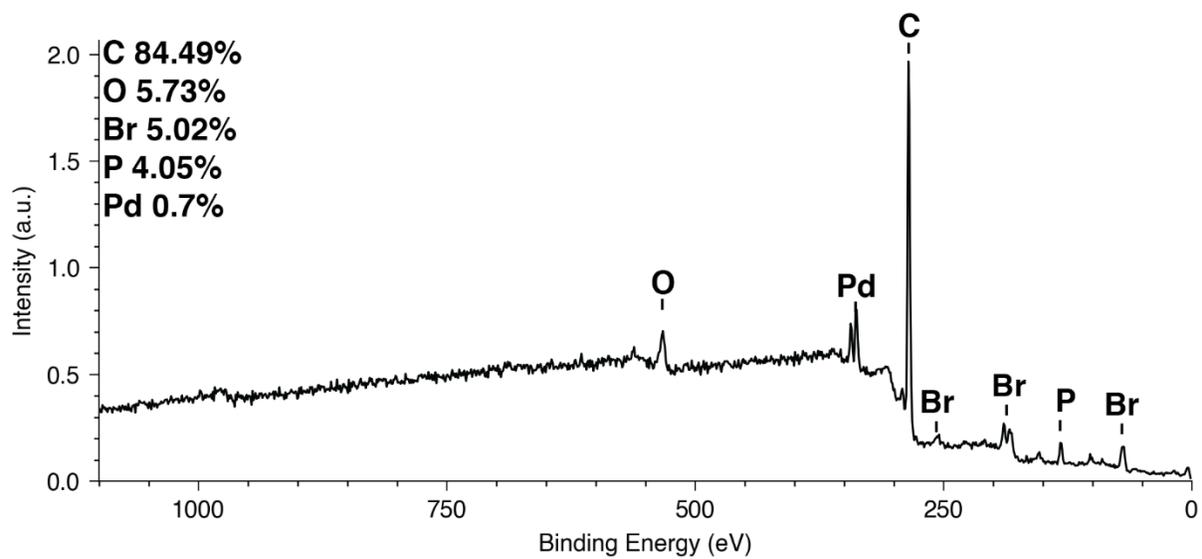

**Fig. S10.** XPS survey corresponding to Table S1, Entry 2 (TBTEB and Pd(PPh$_3$)$_4$/Cu foil in pyridine).

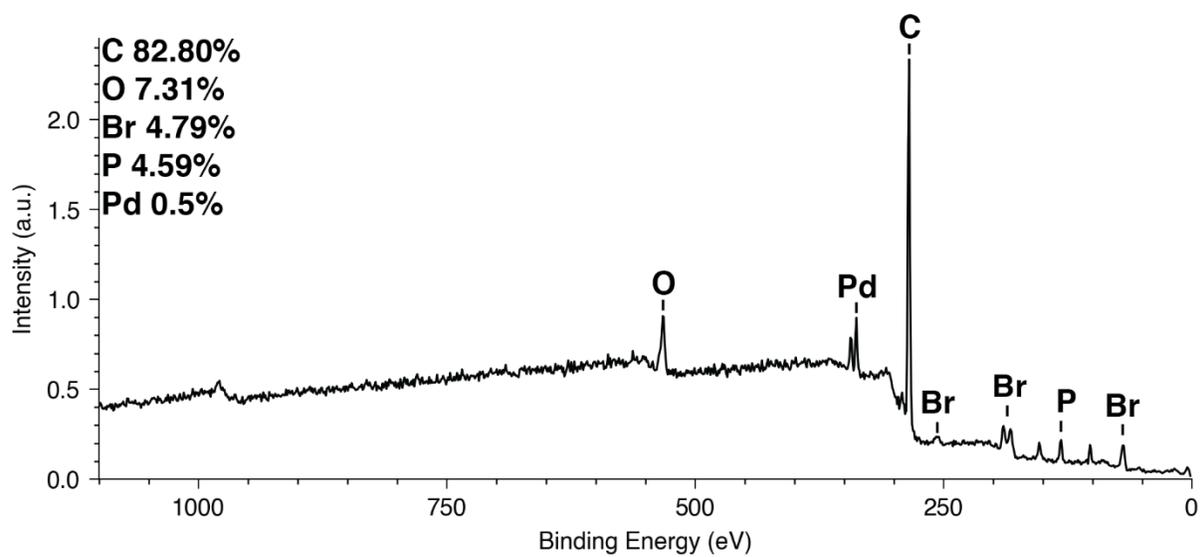

**Fig. S11.** XPS survey corresponding to Table S1, Entry 3 (TBTEB and Pd(PPh$_3$)$_4$ in pyridine, no Cu).

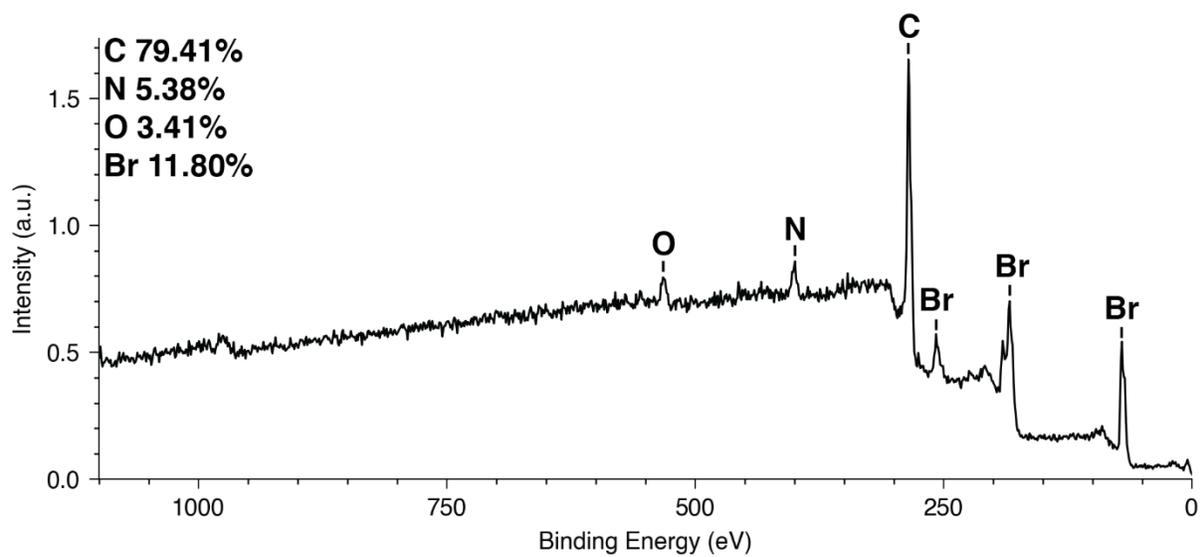

**Fig. S12.** XPS survey corresponding to Table S1, Entry 4 (thermal decomposition of TBTEB in refluxing pyridine).

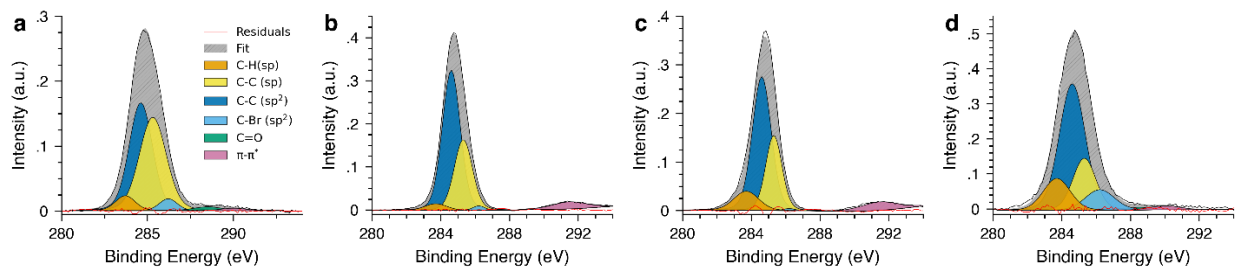

**Fig. S13.** High resolution XPS data for C1s peak regions. **a,** Table S1, Entry 1 (TBTEB and Pd(PPh$_3$)$_4$/CuI in pyridine). **b,** Table S1, Entry 2 (TBTEB and Pd(PPh$_3$)$_4$/Cu foil in pyridine). **c,** Table S1, Entry 3 (control experiment with TBTEB and Pd(PPh$_3$)$_4$ in pyridine, no Cu). **d,** Table S1, Entry 4 (control experiment with no catalysts).

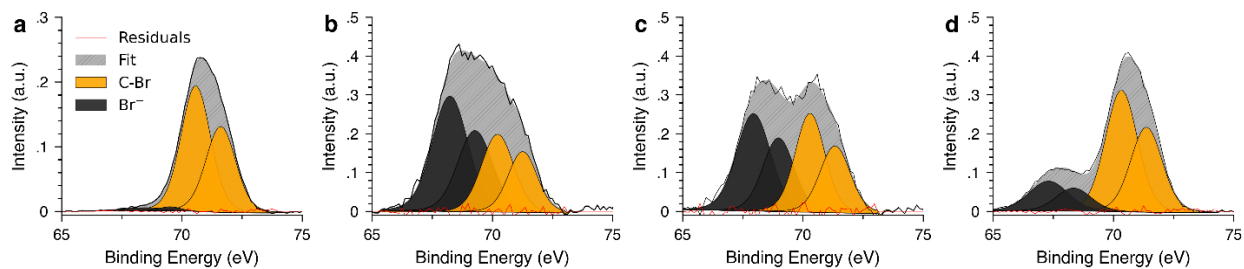

**Fig. S14.** High resolution Br 3d region XPS spectra of selected carbon material samples. **a,** Table S1, Entry 1 (TBTEB and Pd(PPh$_3$)$_4$/CuI in pyridine). **b,** Table S1, Entry 2 (TBTEB and Pd(PPh$_3$)$_4$/Cu foil in pyridine). **c,** Table S1, Entry 3 (control experiment with TBTEB and Pd(PPh$_3$)$_4$ in pyridine, no Cu). **d,** Table S1, Entry 4 (control experiment with no catalysts).

High Resolution Br 3d XPS

High-resolution Br 3d XPS spectra of the carbon materials from Table S1, Entries 1-4 are presented in Fig. S14. The Br signal manifests as a pair of peaks due to spin-orbit coupling, yielding a Br$3d_{3/2}$ peak at a higher binding energy and Br$3d_{5/2}$ peak at a lower value. These coupled peaks are well-defined in XPS, with a separation of 1.04 eV and intensity ratio of 0.671.

Prior to the reaction, bromine species constitute ~20% atomic percent (at%) of the monomer. A decrease in this fraction is observed in all cases after the reaction. The residual Br 3d peak can be deconvolved to two distinctive species: Br covalently bonded to aromatic carbon (71.4, 70.5 eV),[7, 41, 42] corresponding to partially unreacted sites, and weakly coordinated/anionic Br (67.5-69.5 eV)[42-46] trapped within the carbon matrix or on the edges or surfaces of γ-graphyne sheets.

After the reaction catalyzed by Pd(PPh$_3$)$_4$/CuI in pyridine (Table S1, Entry 1) XPS survey indicates an apparent 12 at% residual Br (Fig. S9), consisting almost entirely of covalent C(Ar)-Br (Fig. S14a). The structure of the fit includes contributions from at least two C(Ar)-Br species, presumably mono- and di-substituted. As this material consists of micron-scale crystallites (Fig. S16a-d), and the surface penetration depth for XPS in carbon materials is 10 nm or less, we attribute this C(Ar)-Br signal to the unreacted edge groups. Because of the composition difference between edge and bulk of the material, a survey XPS measurement likely results in a significant overestimation of the Br content. It is important to note that the number of terminal alkyne C-H groups estimated from fitting of the C1s XPS signal (Fig. 2d, Main Text and S13a) correlates well with the apparent C(Ar)-Br content, as would be expected for complementary edge groups.

For the rest of the reactions surveyed (Table S1, Entries 2-4) the carbon products are significantly contaminated with the weakly bound Br species (Fig. S14b-d). As these products are significantly less crystalline than the product of the optimized Pd(PPh$_3$)$_4$/CuI protocol, small-molecule impurities get trapped in disordered carbon matrices. This is also evidenced through observation of Pd, P, and N contaminants derived from solvents and/or catalysts in the corresponding survey XPS spectra (Figs. S9-S12). While the small molecule contamination precludes quantitative interpretation of the results, important chemical insights can be obtained by examining the speciation of Br in the products. It can be clearly seen that thermal decomposition of TBTEB is accompanied by Br loss, likely through spontaneous hydrodebromination (Fig S14d). Pd(PPh$_3$)$_4$ alone is capable of activating the C(Ar)-Br sites, accelerating the rate of hydrodebromination (compare Fig. S14c and S14d).

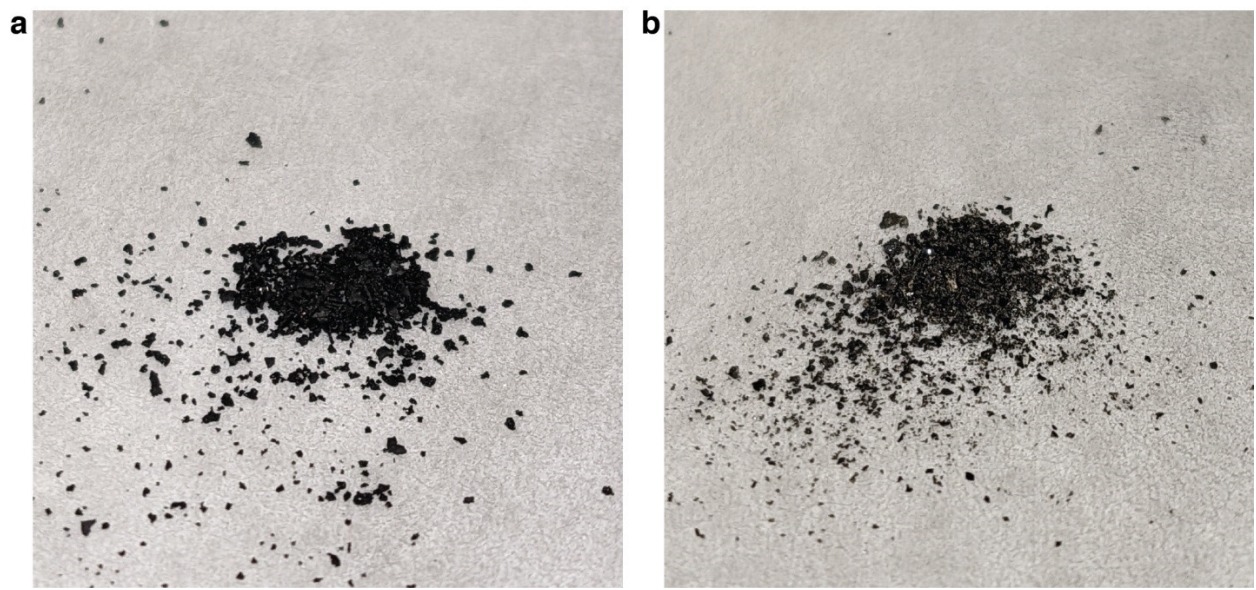

**Fig. S15. a,** Photo of the carbon material corresponding to Table S1, Entry 1 (TBTEB and Pd(PPh$_3$)$_4$/CuI in pyridine). **b,** Photo of the carbon material corresponding to Table S1, Entry 2 (TBTEB and Pd(PPh$_3$)$_4$/Cu foil in refluxing pyridine).

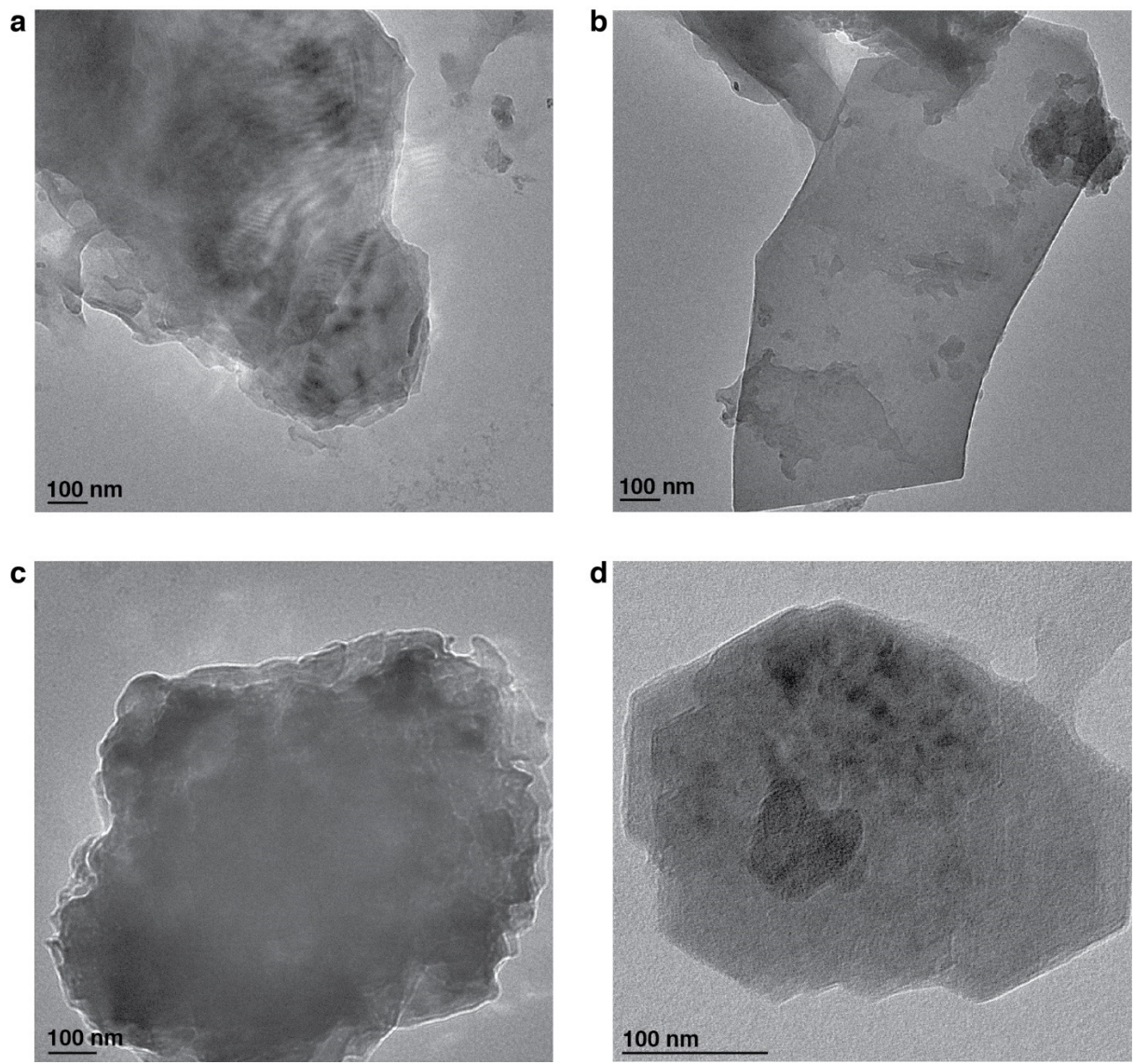

**Fig. S16. a-d,** Representative bright field TEM images of the carbon product corresponding to Table S1, Entry 1 (TBTEB and Pd(PPh$_3$)$_4$/CuI in pyridine).

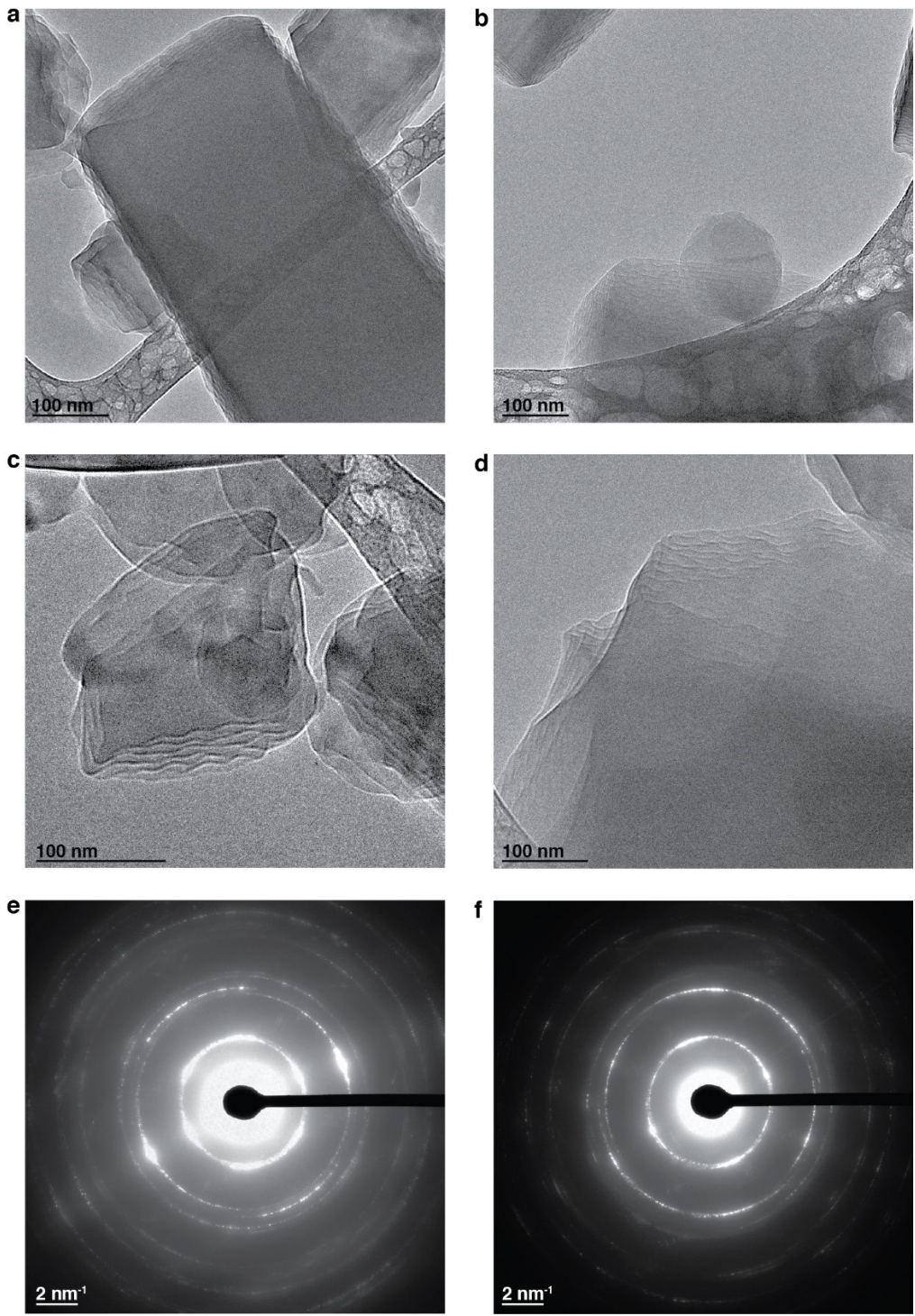

**Fig. S17. a-f,** Representative bright field TEM images of the carbon product corresponding to Table S1, Entry 2 (TBTEB and Pd(PPh$_3$)$_4$/Cu foil in pyridine). (**e-f**) Representative SAED ring patterns for the same material.

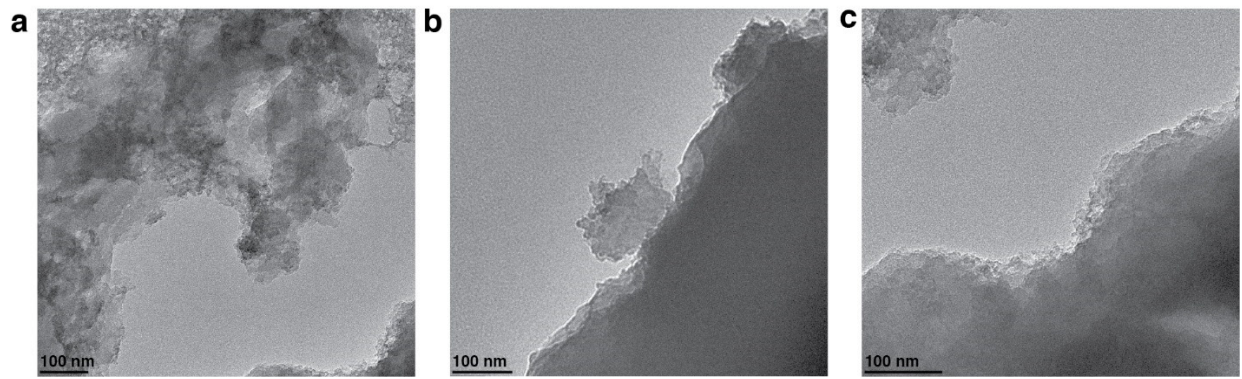

**Fig. S18. a-c,** Representative bright field TEM images of the carbon product corresponding to Table S1, Entry 3 (control experiment with TBTEB and Pd(PPh$_3$)$_4$ in pyridine, no Cu).

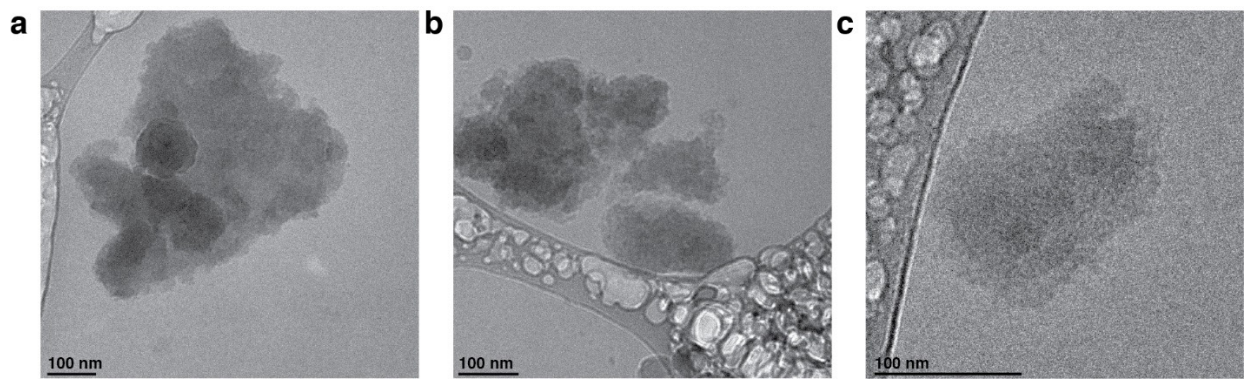

**Fig. S19. a-c,** Representative bright field TEM images of the carbon product corresponding to Table S1, Entry 4 (thermal decomposition of TBTEB in refluxing pyridine).

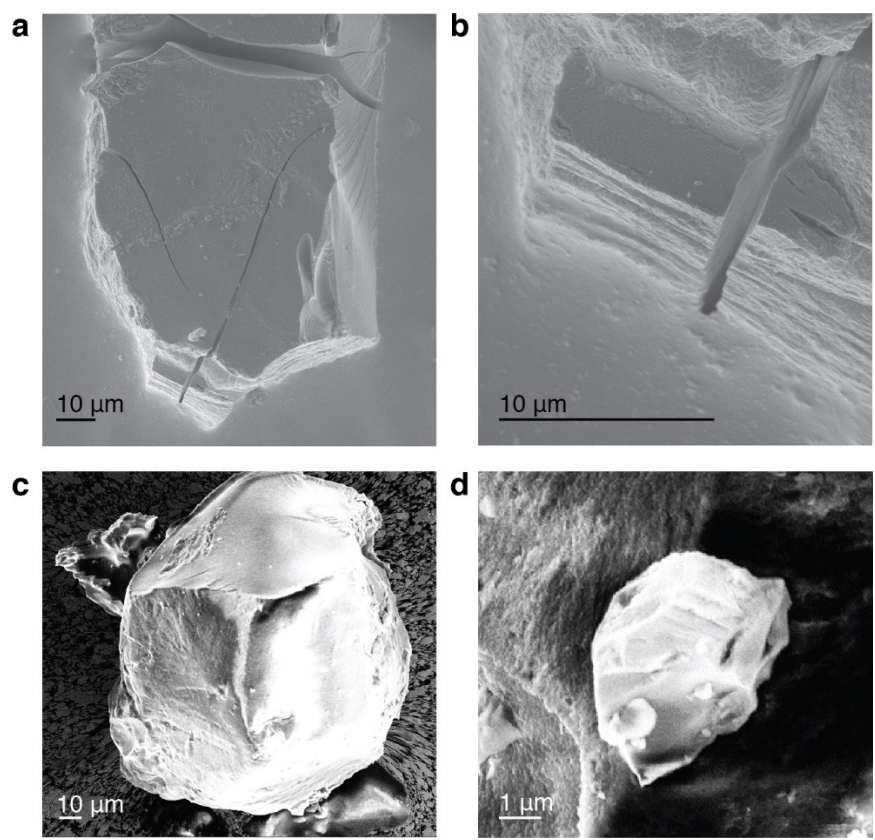

**Fig. S20. a-d**, Representative SEM images of the carbon product corresponding to Table S1, Entry 1 (TBTEB and Pd(PPh$_3$)$_4$/CuI in pyridine).

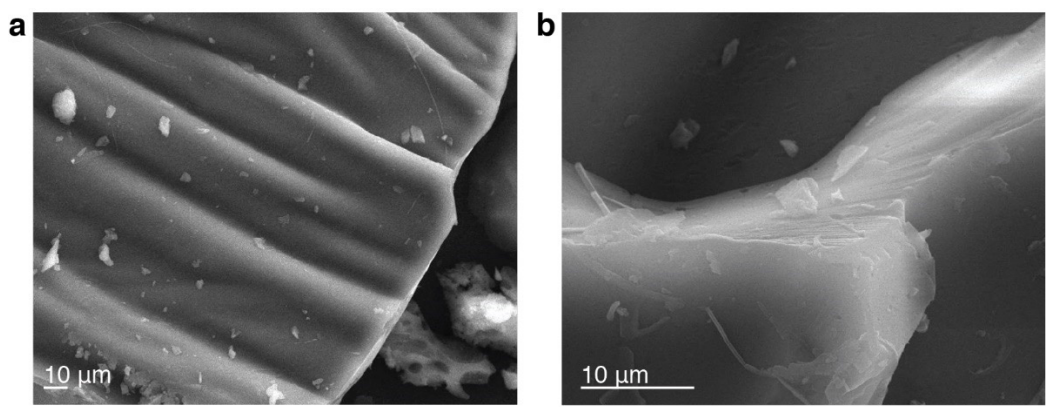

**Fig. S21. a-b**, Representative SEM images of the carbon product corresponding to Table S1, Entry 2 (TBTEB and Pd(PPh$_3$)$_4$/Cu foil in pyridine).

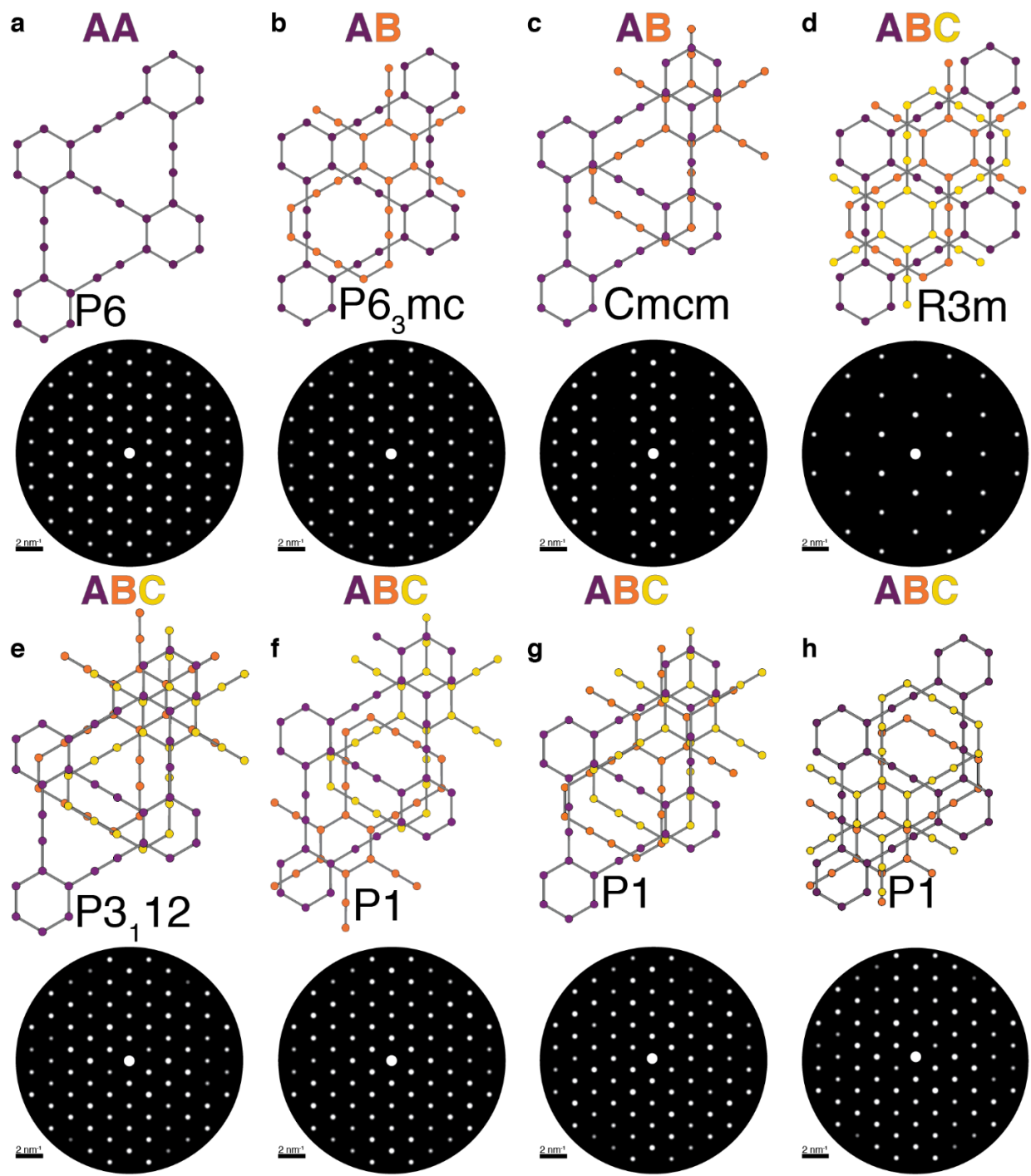

**Fig. S22.** Possible stacking modes of γ-graphyne sheets (**a-h**, top) and their simulated SAED patterns in the *c* orientation (sheets perpendicular to the incident beam, **a-h**, bottom). Interatomic distances are obtained from DFT calculations.

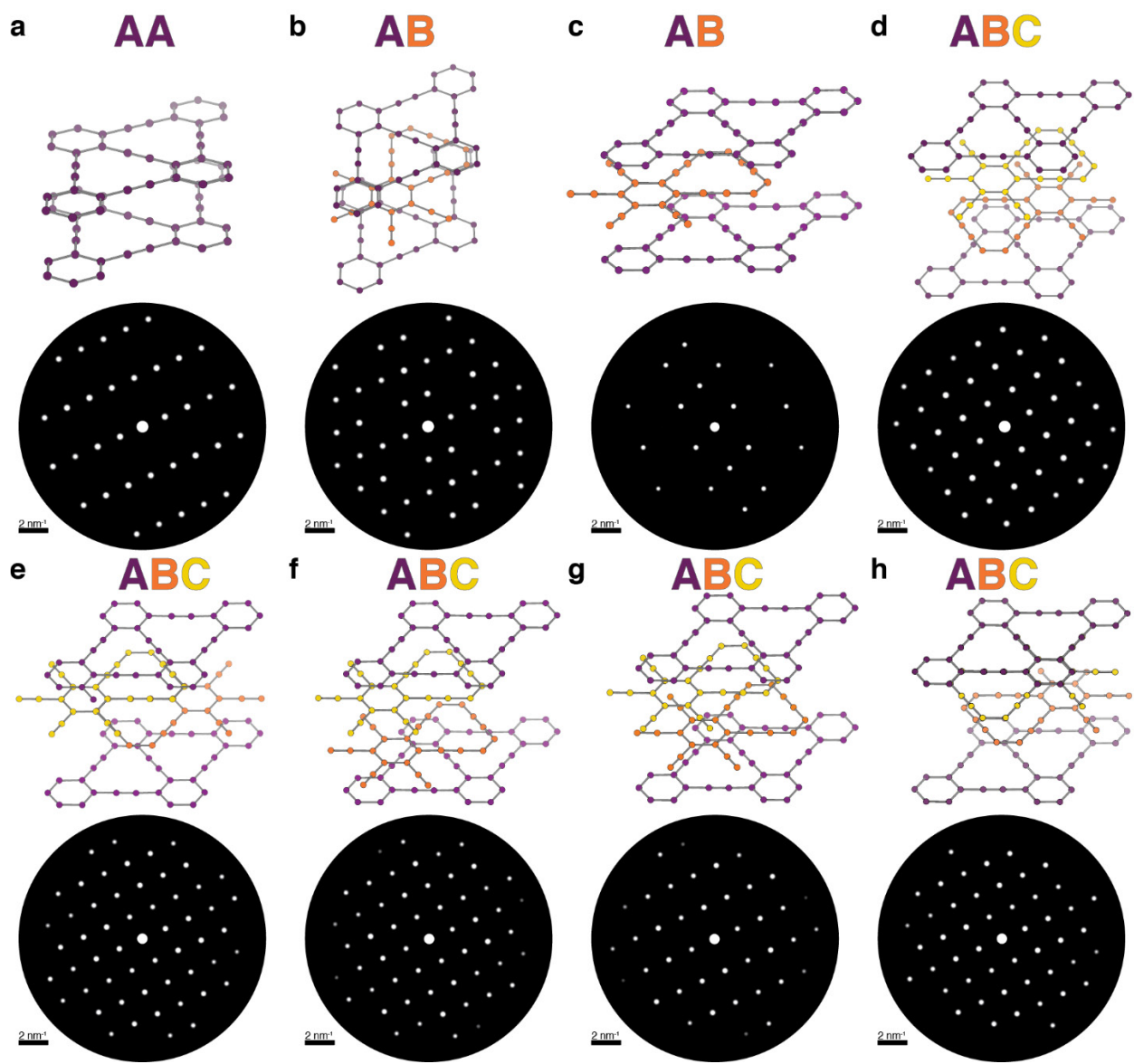

**Fig. S23.** Possible stacking modes of γ-graphyne sheets (**a-h**, top) and their simulated SAED patterns for 45° rotation from the (0001) pole to a viewing direction (matching the experimental diffraction pattern rotated 45°, **a-h**, bottom). Interatomic distances are obtained from DFT calculations.

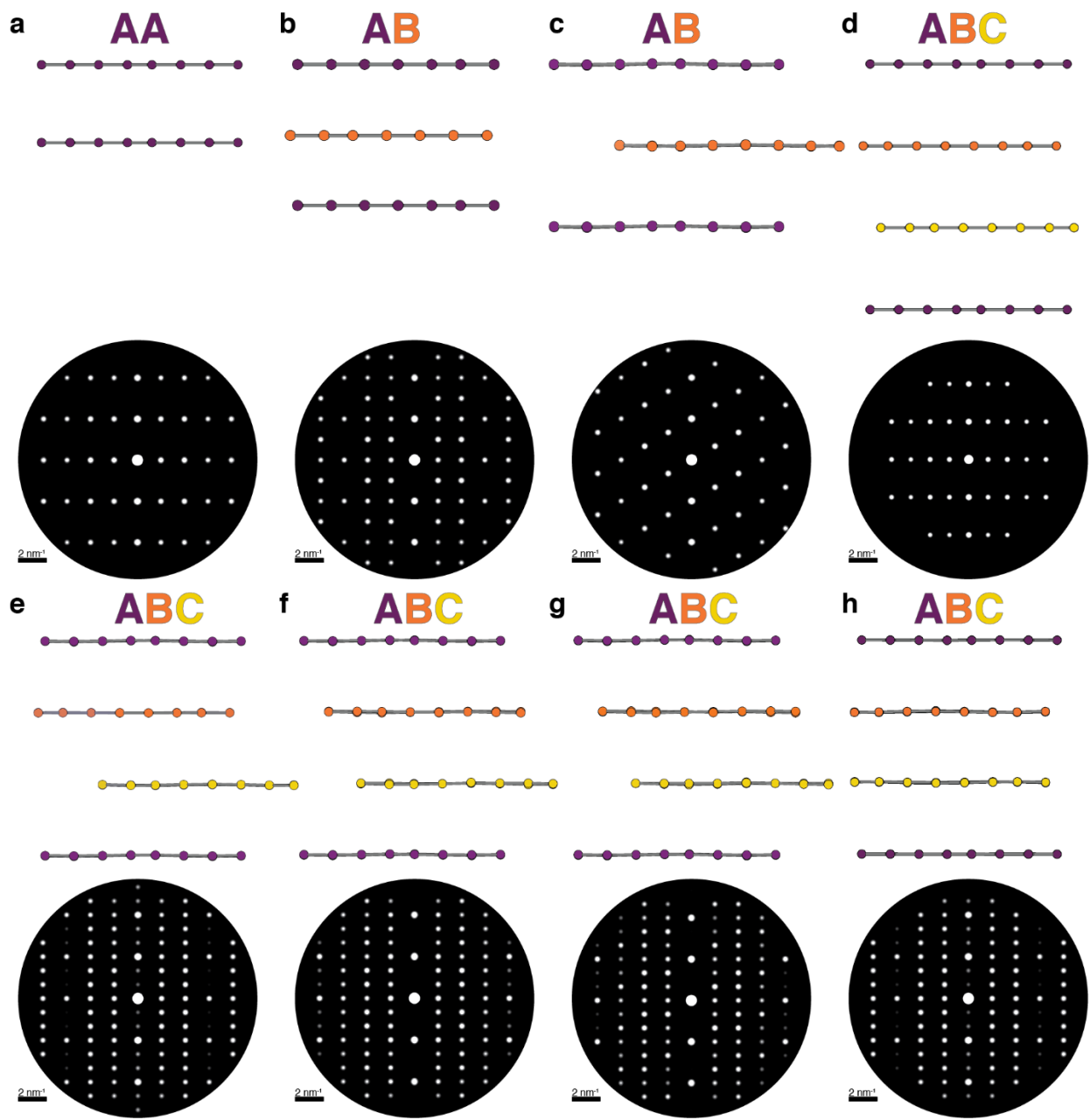

**Fig. S24.** Possible stacking modes of γ-graphyne sheets (**a-h**, top) and their simulated SAED patterns in the *b* orientation (sheets parallel to the incident beam, **a-h**, bottom). Interatomic distances are obtained from DFT calculations.

**Table S2.** Crystallographic information for the structure in Fig 22-24a.

| AA Stacking  File: AA.cif | | | |
|---|---|---|---|
| Hexagonal  P 6  a = b = 6.8826 Å c = 3.4 Å  α = β = 90°, γ = 120° | | | |
| Atom | X | Y | Z |
| C1 | 0.205 | 0 | 0 |
| C2 | 0.412 | 0 | 0 |

**Table S3.** Crystallographic information for the structure in Fig 22-24b.

| AB Stacking<br>File: AB_1_P_63_mmc.cif | | |
|---|---|---|
| Hexagonal<br>P 6$_3$ m c<br>a = b = 6.8826 Å  c = 6.8 Å<br>α = β = 90°, γ = 120° | | |
| Atom | X | Y | Z |
| C1 | 0.12833 | 0.66667 | 0 |
| C2 | 0.9213 | 0.66667 | 0 |

**Table S4.** Crystallographic information for structure in Fig 22-24c.

| AB Stacking  File: AB_2_C_mcm.cif |||
|---|---|---|
| Orthorhombic  C m c m  a = 11.86455 b = 6.8826 Å c = 6.8 Å  α = β = γ = 90° |||
| Atom | X | Y | Z |
| C1 | 0.60351 | 0.27153 | 0.25 |
| C2 | 0.7056 | 0.16932 | 0.25 |
| C3 | 0 | 0.08195 | 0.25 |
| C4 | 0 | 0.28643 | 0.25 |
| C5 | 0.39653 | 0.47845 | 0.25 |
| C6 | 0.79423 | 0.08036 | 0.25 |
| C7 | 0 | 0.66825 | 0.25 |
| C8 | 0 | 0.46396 | 0.25 |

**Table S5.** Crystallographic information for the structure in Fig 22-24d.

| ABC Stacking File: ABC_1_R_3m.cif | | | |
|---|---|---|---|
| Trigonal R 3 m a = 6.8826 Å  b = 6.8826 Å  c = 10.2 Å α = β = 90°, γ = 120° | | | |
| Atom | X | Y | Z |
| C1 | 0.205 | 0 | 0 |
| C2 | 0.412 | 0 | 0 |

**Table S6.** Crystallographic information for the structure in Fig 22-24e.

| ABC Stacking<br>File: ABC_2_P_31_1_2.cif | | | |
|---|---|---|---|
| Trigonal<br>P3$_1$ 1 2<br>a = 6.8826 Å  b = 6.8826 Å  c = 10.2 Å<br>α = β = 90°, γ = 120° | | | |
| Atom | X | Y | Z |
| C1 | -0.0461 | 0.25305 | 0.33416 |
| C2 | 0.15811 | 0.25303 | 0.33675 |
| C3 | 0.74697 | 0.45999 | 0.33221 |
| C4 | 0.74719 | 0.66451 | 0.33033 |
| C5 | 0.54 | 0.04594 | 0.33418 |
| C6 | 0.33594 | 0.84148 | 0.33416 |

**Table S7.** Crystallographic information for the structure in Fig 22-24f.

| ABC Stacking File: ABC_3_P_1.cif |||
|---|---|---|---|
| Triclinic P 1 a = 6.8826 Å b = 6.8826 Å c = 10.2 Å α = β = 90°, γ = 120° ||||
| Atom | X | Y | Z |
| C1 | 0.2082 | 0.00369 | 0.03346 |
| C2 | 0.41232 | 0.00348 | 0.03615 |
| C3 | 0.00125 | 0.21104 | 0.02974 |
| C4 | 0.00128 | 0.41553 | 0.02911 |
| C5 | 0.7941 | 0.7971 | 0.03339 |
| C6 | 0.58989 | 0.59289 | 0.036 |
| C7 | 0.79437 | 0.00396 | 0.03178 |
| C8 | 0.58975 | 0.00353 | 0.03455 |
| C9 | 0.00093 | 0.79713 | 0.03197 |
| C10 | 0.00084 | 0.59282 | 0.03011 |
| C11 | 0.20831 | 0.21084 | 0.03177 |
| C12 | 0.41291 | 0.41502 | 0.03447 |
| C13 | 0.20746 | 0.25244 | 0.36408 |
| C14 | 0.41183 | 0.25267 | 0.3614 |
| C15 | 0.00061 | 0.45964 | 0.36408 |
| C16 | 0.00071 | 0.66389 | 0.36511 |
| C17 | 0.79344 | 0.04594 | 0.36413 |
| C18 | 0.58888 | 0.84196 | 0.36166 |
| C19 | 0.79355 | 0.25266 | 0.36279 |
| C20 | 0.58943 | 0.25262 | 0.36005 |
| C21 | 0.00029 | 0.04573 | 0.36571 |
| C22 | 0.00025 | 0.84136 | 0.3658 |
| C23 | 0.2077 | 0.45976 | 0.36282 |
| C24 | 0.41172 | 0.66394 | 0.36035 |
| C25 | 0.5396 | -0.0693 | 0.6984 |
| C26 | 0.74397 | -0.0694 | 0.69948 |
| C27 | 0.33301 | 0.13814 | 0.6988 |
| C28 | 0.33305 | 0.34252 | 0.69902 |
| C29 | 0.12586 | 0.7244 | 0.69882 |
| C30 | -0.0786 | 0.52011 | 0.69897 |
| C31 | 0.12615 | -0.0688 | 0.69937 |
| C32 | -0.0782 | -0.0686 | 0.70003 |
| C33 | 0.33238 | 0.72407 | 0.69829 |
| C34 | 0.33218 | 0.51958 | 0.69867 |
| C35 | 0.53995 | 0.13784 | 0.69835 |
| C36 | 0.74438 | 0.34213 | 0.69858 |

**Table S8.** Crystallographic information for the structure in Fig 22-24g.

| ABC Stacking File: ABC_4_P_1.cif | | | |
|---|---|---|---|
| Triclinic P 1 a = 6.8826 Å b = 6.8826 Å c = 10.2 Å $\alpha = \beta = 90°, \gamma = 120°$ | | | |
| Atom | X | Y | Z |
| C1 | 0.20667 | 0.00525 | 0.03487 |
| C2 | 0.99656 | 0.04592 | 0.37042 |
| C3 | 0.99663 | 0.84175 | 0.37282 |
| C4 | 0.20381 | 0.46018 | 0.36719 |
| C5 | 0.40783 | 0.66417 | 0.36718 |
| C6 | 0.96727 | 0.26154 | 0.70381 |
| C7 | 0.1718 | 0.26159 | 0.70446 |
| C8 | 0.5854 | 0.25291 | 0.36175 |
| C9 | 0.76012 | 0.46863 | 0.69988 |
| C10 | 0.55331 | 0.05453 | 0.69973 |
| C11 | 0.34904 | 0.85037 | 0.697 |
| C12 | 0.55351 | 0.26145 | 0.70105 |
| C13 | 0.34928 | 0.26152 | 0.70302 |
| C14 | 0.76017 | 0.0547 | 0.70185 |
| C15 | 0.75975 | 0.85009 | 0.69941 |
| C16 | 0.75989 | 0.67277 | 0.69767 |
| C17 | 0.78956 | 0.25282 | 0.36422 |
| C18 | 0.58537 | 0.84163 | 0.36731 |
| C19 | 0.78967 | 0.04595 | 0.36734 |
| C20 | 0.41085 | 0.00506 | 0.03731 |
| C21 | 0.99956 | 0.21238 | 0.03077 |
| C22 | 0.99958 | 0.41685 | 0.02999 |
| C23 | 0.79264 | 0.79833 | 0.03505 |
| C24 | 0.58845 | 0.59413 | 0.03776 |
| C25 | 0.79274 | 0.00519 | 0.03296 |
| C26 | 0.58814 | 0.00481 | 0.03561 |
| C27 | 0.99955 | 0.79858 | 0.03354 |
| C28 | 0.99956 | 0.59434 | 0.03143 |
| C29 | 0.20664 | 0.21233 | 0.03304 |
| C30 | 0.41119 | 0.41657 | 0.03596 |
| C31 | 0.20355 | 0.25287 | 0.36769 |
| C32 | 0.40794 | 0.25318 | 0.36406 |
| C33 | 0.99653 | 0.45992 | 0.36686 |
| C34 | 0.99683 | 0.66426 | 0.37044 |
| C35 | 0.96722 | 0.46863 | 0.70158 |
| C36 | 0.17143 | 0.6732 | 0.69867 |

**Table S9.** Crystallographic information for the structure in Fig 22-24h.

| ABC Stacking File: ABC_5_P_1.cif |||
|---|---|---|
| Triclinic P 1 a = 6.8826 Å  b = 6.8826 Å  c = 10.2 Å α = β = 90°, γ = 120° |||
| Atom | X | Y | Z |

| Atom | X | Y | Z |
|---|---|---|---|
| C1 | 0.20729 | 0.00285 | 0.03076 |
| C2 | 0.4116 | 0.00274 | 0.03013 |
| C3 | 0.0004 | 0.21004 | 0.03198 |
| C4 | 0.00035 | 0.41442 | 0.03159 |
| C5 | 0.79356 | 0.79639 | 0.03191 |
| C6 | 0.58913 | 0.59204 | 0.03168 |
| C7 | 0.7936 | 0.00297 | 0.03192 |
| C8 | 0.58918 | 0.00287 | 0.0308 |
| C9 | 0.0003 | 0.79627 | 0.03138 |
| C10 | 0.00036 | 0.59192 | 0.03118 |
| C11 | 0.20737 | 0.21 | 0.03135 |
| C12 | 0.41183 | 0.41439 | 0.03123 |
| C13 | 0.53792 | 0.6814 | 0.36681 |
| C14 | 0.74207 | 0.68142 | 0.36949 |
| C15 | 0.33055 | 0.88835 | 0.36378 |
| C16 | 0.33035 | 0.09262 | 0.3637 |
| C17 | 0.12385 | 0.47443 | 0.36693 |
| C18 | -0.0802 | 0.27021 | 0.36939 |
| C19 | 0.1238 | 0.68116 | 0.36551 |
| C20 | -0.0807 | 0.6808 | 0.36811 |
| C21 | 0.33076 | 0.47473 | 0.36565 |
| C22 | 0.33079 | 0.27053 | 0.36458 |
| C23 | 0.53784 | 0.88843 | 0.36541 |
| C24 | 0.74236 | 0.09251 | 0.36793 |
| C25 | 0.53617 | -0.0708 | 0.69795 |
| C26 | 0.74074 | -0.0704 | 0.69524 |
| C27 | 0.32902 | 0.13614 | 0.69763 |
| C28 | 0.32888 | 0.34036 | 0.69964 |
| C29 | 0.1223 | 0.72255 | 0.69796 |
| C30 | -0.0823 | 0.51833 | 0.69522 |
| C31 | 0.12216 | -0.0709 | 0.69619 |
| C32 | -0.082 | -0.0708 | 0.69357 |
| C33 | 0.32914 | 0.7225 | 0.70012 |
| C34 | 0.3292 | 0.51803 | 0.70085 |
| C35 | 0.53628 | 0.13642 | 0.69616 |
| C36 | 0.74048 | 0.34073 | 0.69353 |

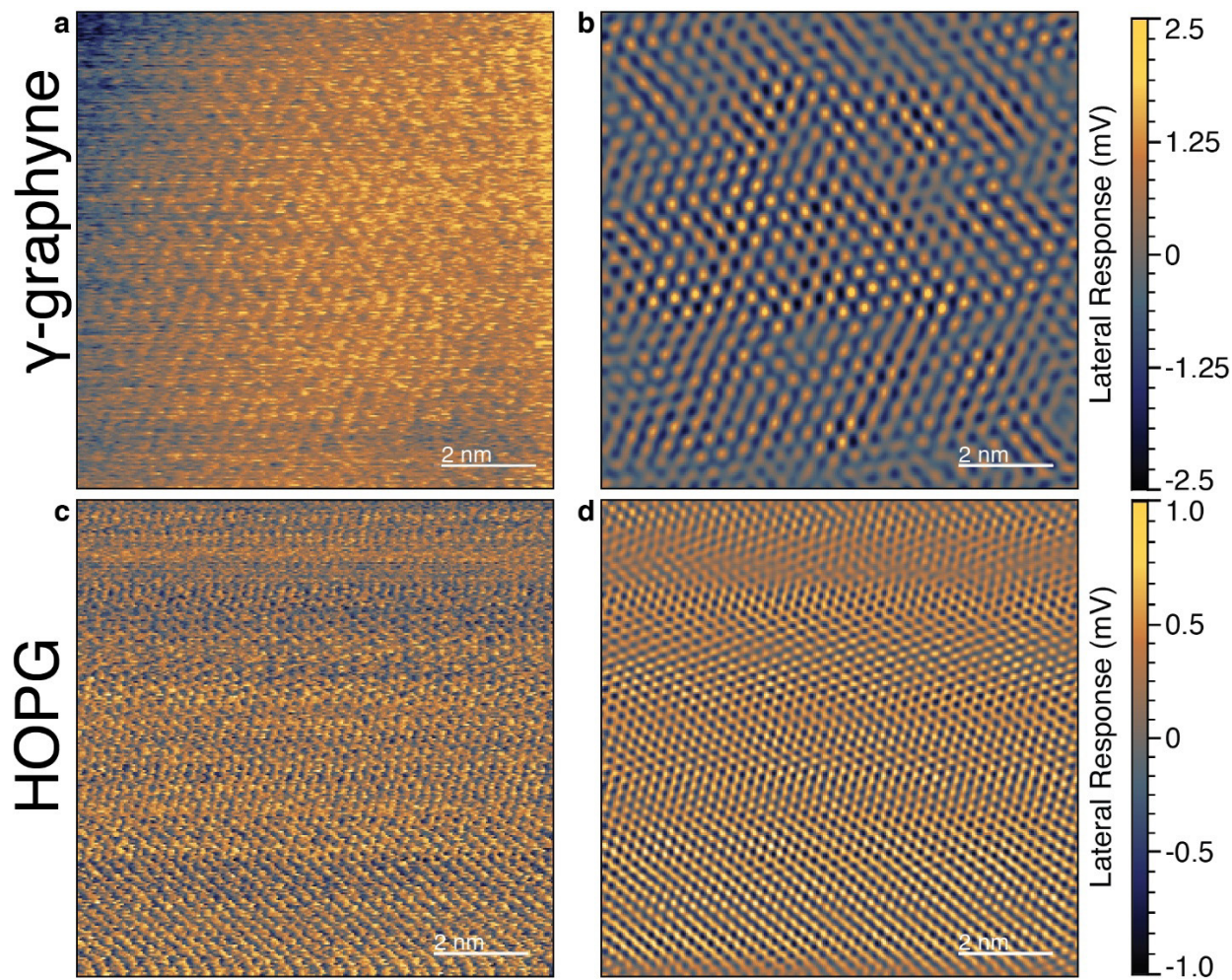

**Fig. S25.** LFM lattice images of the carbon product corresponding to Table S1, Entry 1 (γ-graphyne) and reference HOPG substrate. **a,** Raw LFM image of a 10x10 nm flat area of a γ-graphyne flake (Fig. 4a, Main Text). **b,** Low-pass filtered version of the LFM image in (a). **c,** Raw LFM image of a 10x10 nm area of the reference HOPG substrate obtained several nm from the edge of the sample flake. **d,** Low-pass filtered version of the LFM image in (c).

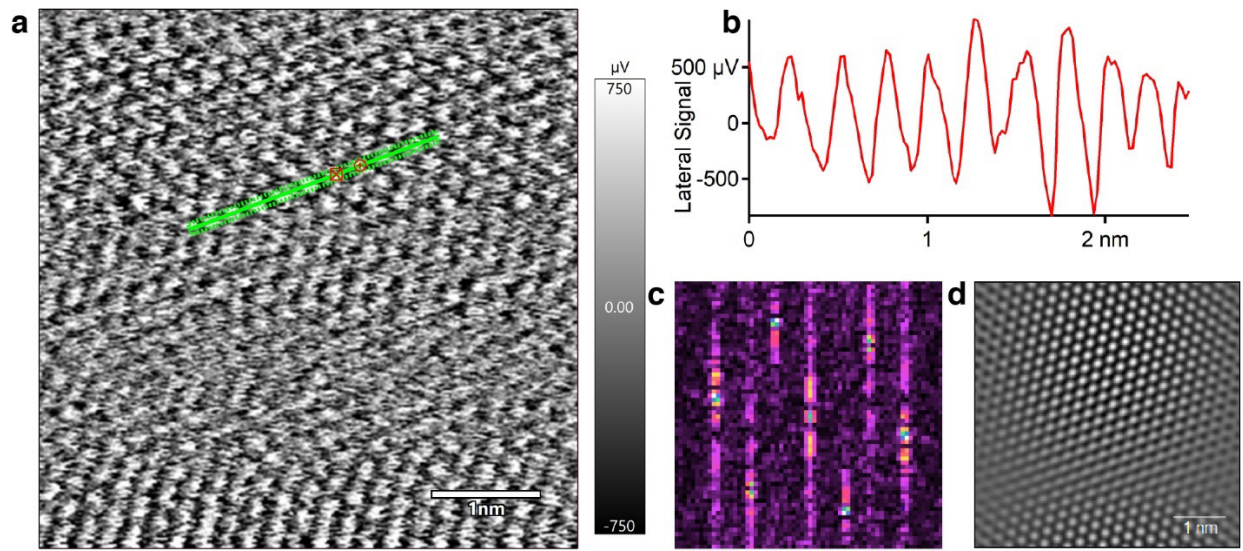

**Fig. S26.** HOPG substrate measurements used for the calibration of the LFM images. **a,** Lateral deflection map of the HOPG surface, raw image. **b,** Plot of lateral response along the linear trace from (a). **c,** FFT of the lateral response map from **a**. **d,** Low-pass filtered lateral deflection map from (a).

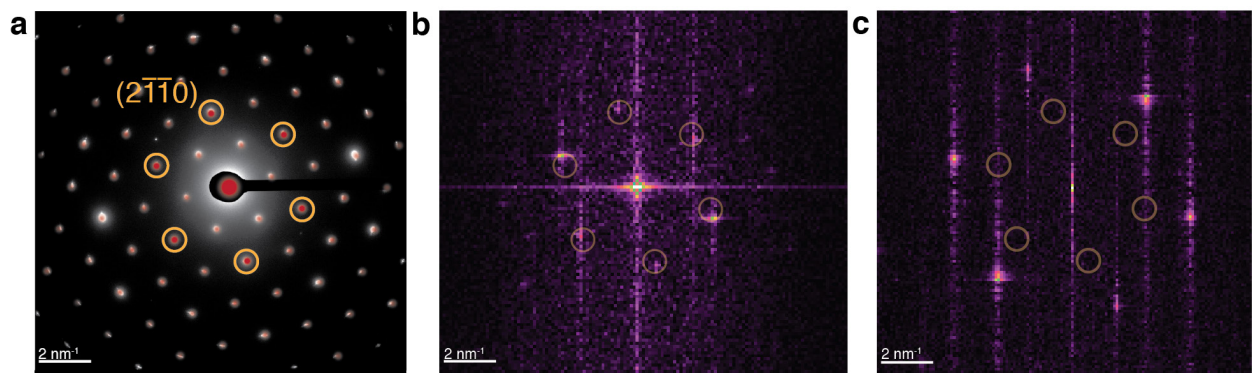

**Fig. S27.** Analysis of periodicity through FFT of scanning microscopy images. **a,** SAED of the product corresponding to Table S1, Entry 1 (γ-graphyne). Reflections corresponding to the $(2\bar{1}\bar{1}0)$ crystallographic planes are highlighted. **b,** FFT of the AFM height image of the same material (Fig. 4a, Main Text). The bright-spot pattern is a perfect match for the reflections highlighted in **a**. **c,** FFT of the LFM image of HOPG substrate (Fig. S25b). The bright-spot pattern is consistent with the expected periodicity for a graphene sheet (distance of ~2.1 Å between $(10\bar{1}0)$ crystallographic planes). The periodicity of HOPG standard is clearly different and from that of γ-graphyne.

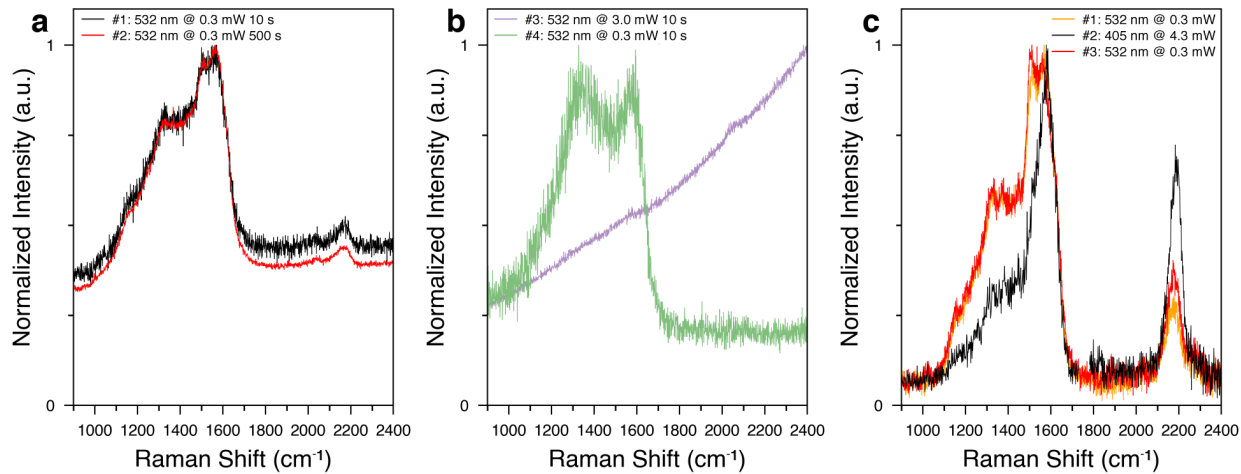

**Fig. S28.** Multi-power and multi-wavelength Raman spectroscopy of γ-graphyne and structural transformations under laser light. **a**, Spectra obtained over 10s (#1, black) and 500s (#2, red) intervals with 532 nm excitation (0.3 mW incident power). The spectra were collected sequentially (10s, then 500s) from the same sample region. **b**, Spectra of the sample region in **a** obtained over 10s at 3.0 mW (magenta, #3) and 0.3 mW (green, #4) incident power. The spectra were obtained sequentially (0.3 mW, then 3.0 mW). The experiment was performed after the low-power experiments in **a**. **c,** Sequentially obtained spectra of γ-graphyne with 532 nm excitation (10s, 0.3 mW incident power, yellow, #1), 405 nm excitation (10s, 4.3 mW incident power, black, #2), and 532 nm excitation (10s, 0.3 mW incident power, red, #3). The spectra were acquired from the same sample region.

Discussion of the Multi-Power Raman Experiments

Figure S28 shows a series of measurements performed to probe whether the transformation of γ-graphyne observed under laser light (Fig. 5b, Main Text) is photochemical or photothermal.

First, a spectrum was obtained over 10s with 532 nm excitation at 0.3 mW incident power (Fig. S28a, black trace). Following this, a spectrum was collected for 500s from the same region at the same power setting (Fig. S28a, red trace). This experiment delivered 50× the photon exposure dose compared to the baseline 10s experiment. No change in the Raman signature was observed.

Then, the incident power was increased to 3.0 mW, and a spectrum of the same sample region was collected over 10s (Fig. S28b, magenta trace). This spectrum is dominated by the high fluorescent background characteristic of disordered/polymeric carbons.[47] Finally, a spectrum of the same sample region is collected over 10s at 0.3 mW incident power (Fig. S28, green trace). This spectrum is similar to the typical spectra of disordered graphitic materials. No γ-graphyne-specific features are discernible, indicating a near-instant structural transformation during the high-power exposure. The photon exposure dose delivered in the 3.0 mW high power experiment is 10× compared to the baseline experiment, but only 20% that of the 500s low-power experiment (Fig. S28a, black trace). Thus, the degree of transformation does not linearly correlate with the exposure dose, indicating thresholding photochemical behavior or a photothermal process.

The experiment in Fig. S28c provides evidence against a purely photothermal process. In this experiment, the 4.3 mW high power 405 nm light exposure does not induce a structural transformation, with stable power 532 nm spectra taken before and after the 405 nm exposure showing no changes. If the process was purely photothermal, then the 4.3 mW 405 nm exposure should have induced transformation similar to the 3.0 mW 532 nm exposure (Fig S29b). Our current hypothesis is that γ-graphyne is photochemically sensitive to 532 nm light above a certain power threshold. However, we cannot conclusively eliminate a photothermal mechanism and speculate that a combination of both photochemical and photothermal transformations is possible.

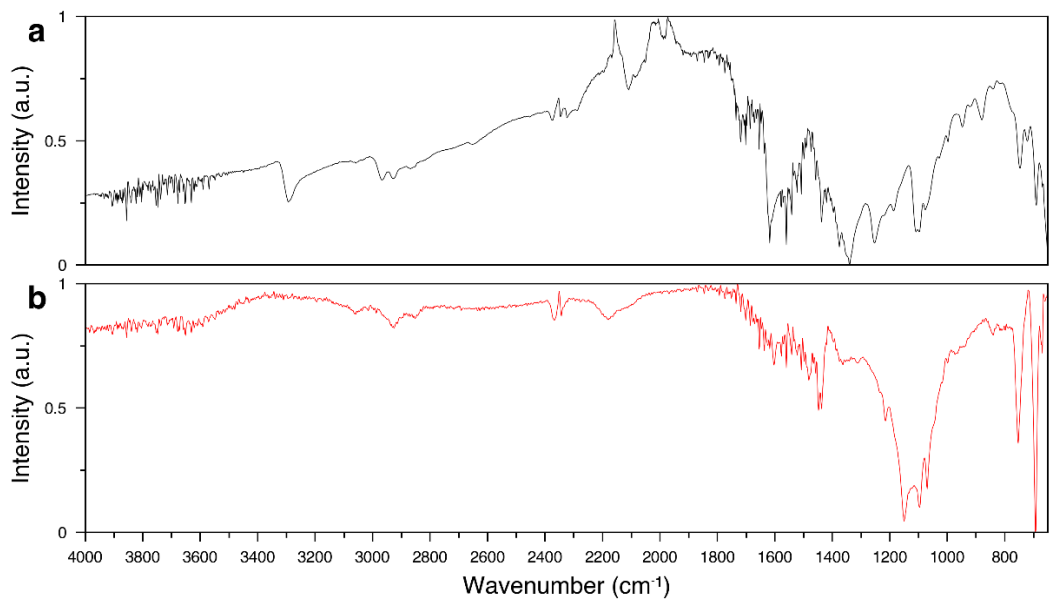

**Fig. S29.** ATR-FTIR spectroscopy of bulk γ-graphyne. **a,** diamond crystal, and **b,** Ge crystal. The prominent feature at ~2100 cm$^{-1}$ in spectrum **a** is a diamond crystal artifact and not the acetylenic absorbance band of γ-graphyne.

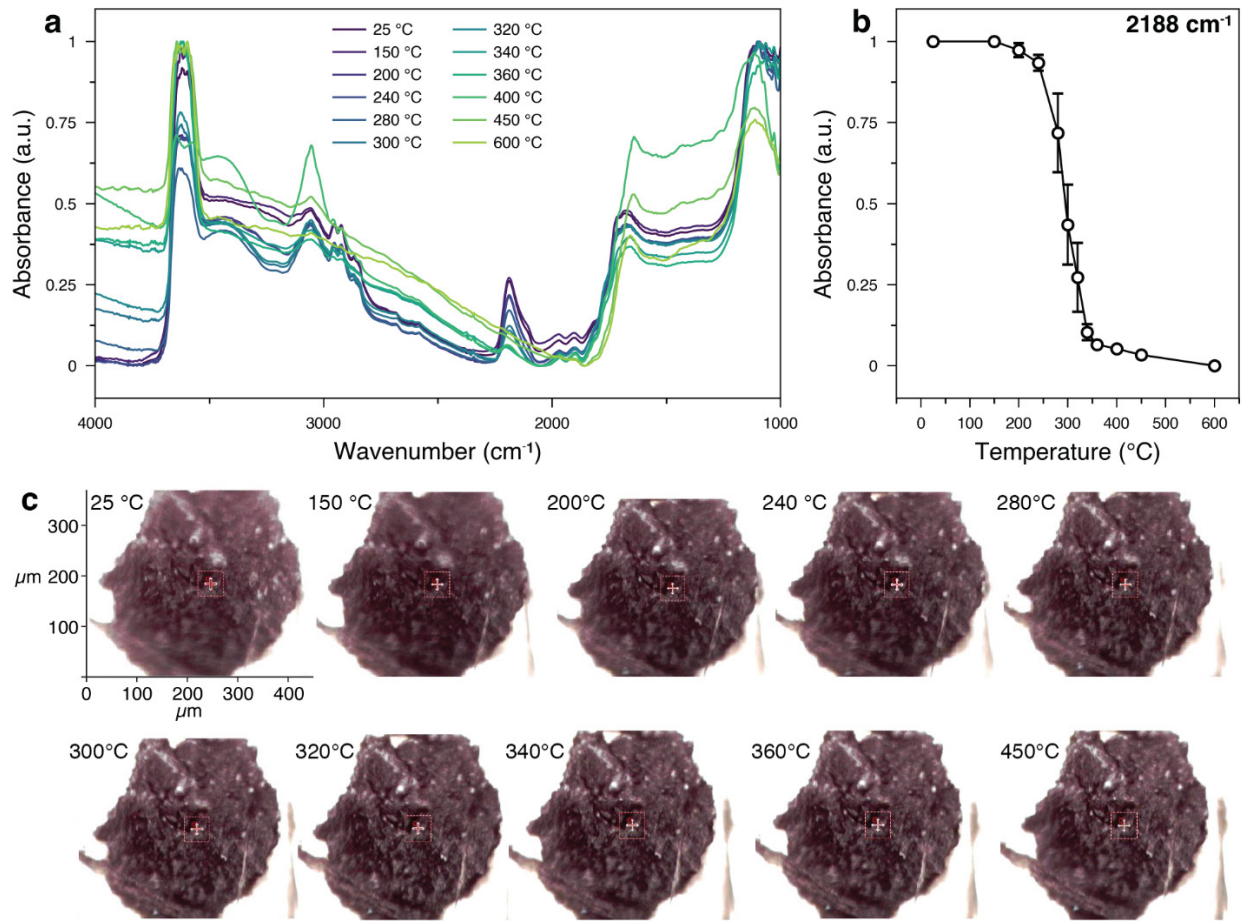

**Fig. S30.** Evaluation of thermal stability of γ-graphyne. **a**, FTIR spectra of a polycrystalline γ-graphyne particle that was heated to increasing set-point temperatures under vacuum (1.3×10$^{-3}$ mbar). The sample was kept at every temperature set-point for one hour, and then cooled down to room temperature for spectroscopic measurements. The indicated temperature is the set-point temperature before the spectral measurement. The sample was suspended in a round hole (D ~200 μm) fabricated in mica film (10x10x0.01mm), and fixed at the edges using bundles from a thin forest-drawn carbon nanotube sheet. **b**, Intensity of the internal alkyne peak at 2188 cm$^{-1}$ as a function of the maximum setpoint temperature for successive annealing. **c**, Picture series for the polycrystalline particle that was heated in vacuum from ambient temperature up to an upper set-point temperature of 450°C, showing no noticeable changes in particle dimensions or texture.

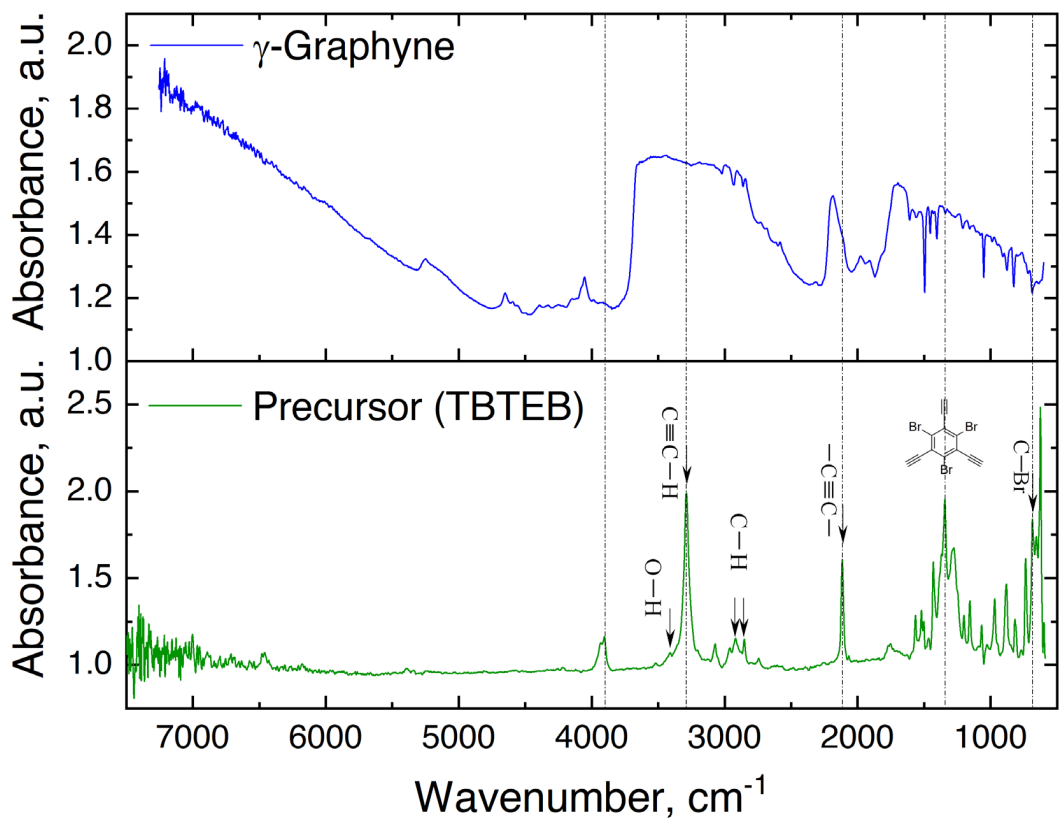

**Fig. S31.** Micro-FTIR spectra of γ-graphyne and TBTEB monomer. Top panel: absorbance spectrum of γ-graphyne for a ~300x300 μm polycrystalline pellet 20 μm thick. Bottom panel: absorbance spectrum of TBTEB. The spectra were collected in reflection mode using a 50x50 μm aperture, 8 cm$^{-1}$ resolution, and 100 scans.

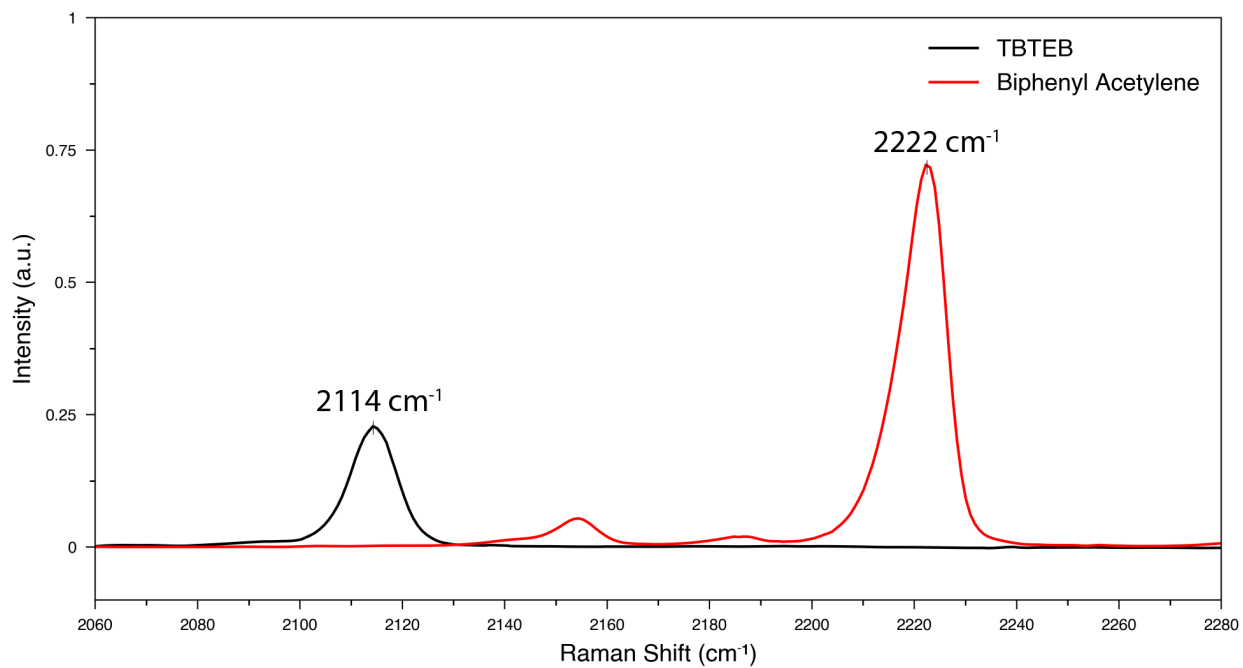

**Fig. S32.** Raman spectra of TBTEB and biphenyl acetylene demonstrating the frequency shift of terminal alkyne compared to internal alkyne.

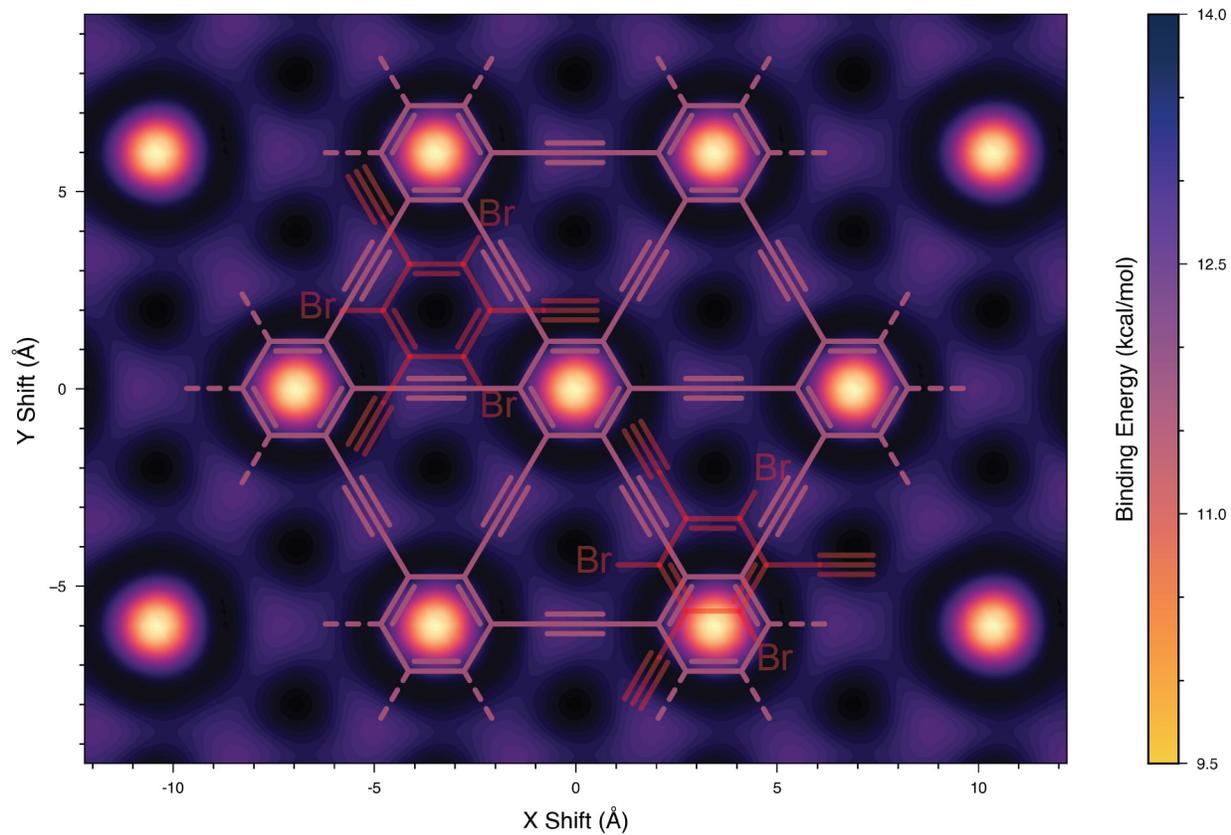

**Fig. S33.** DFT-generated potential energy surfaces for binding of TBTEB monomer to a γ-graphyne monolayer.

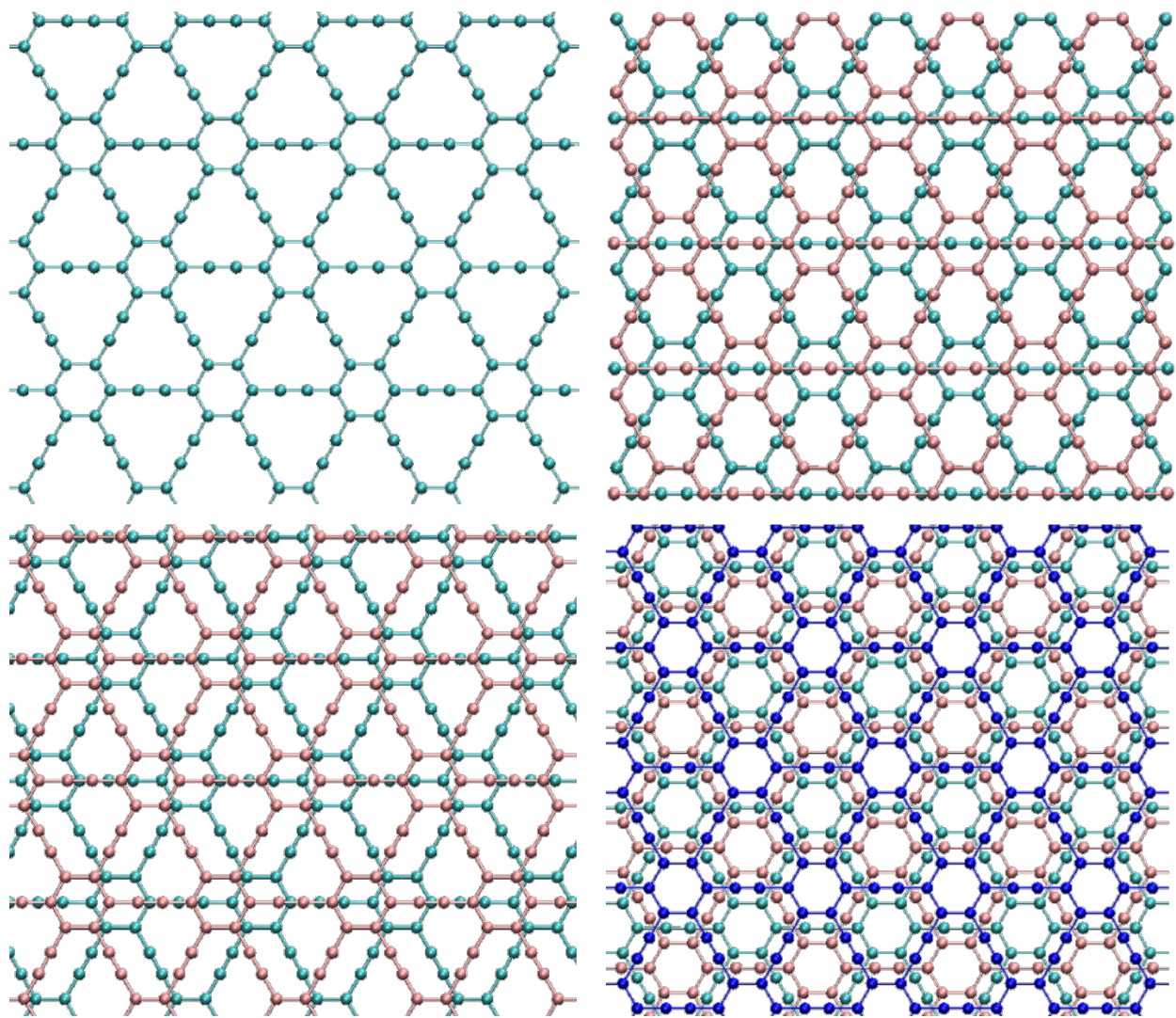

**Fig. S34.** Upper views of the γ-graphyne stacking modes used for molecular dynamics (MD) simulations. From top-left to bottom-right, the packing modes are: AA, AB1, AB2 and ABC. Colors are used to highlight different layers.

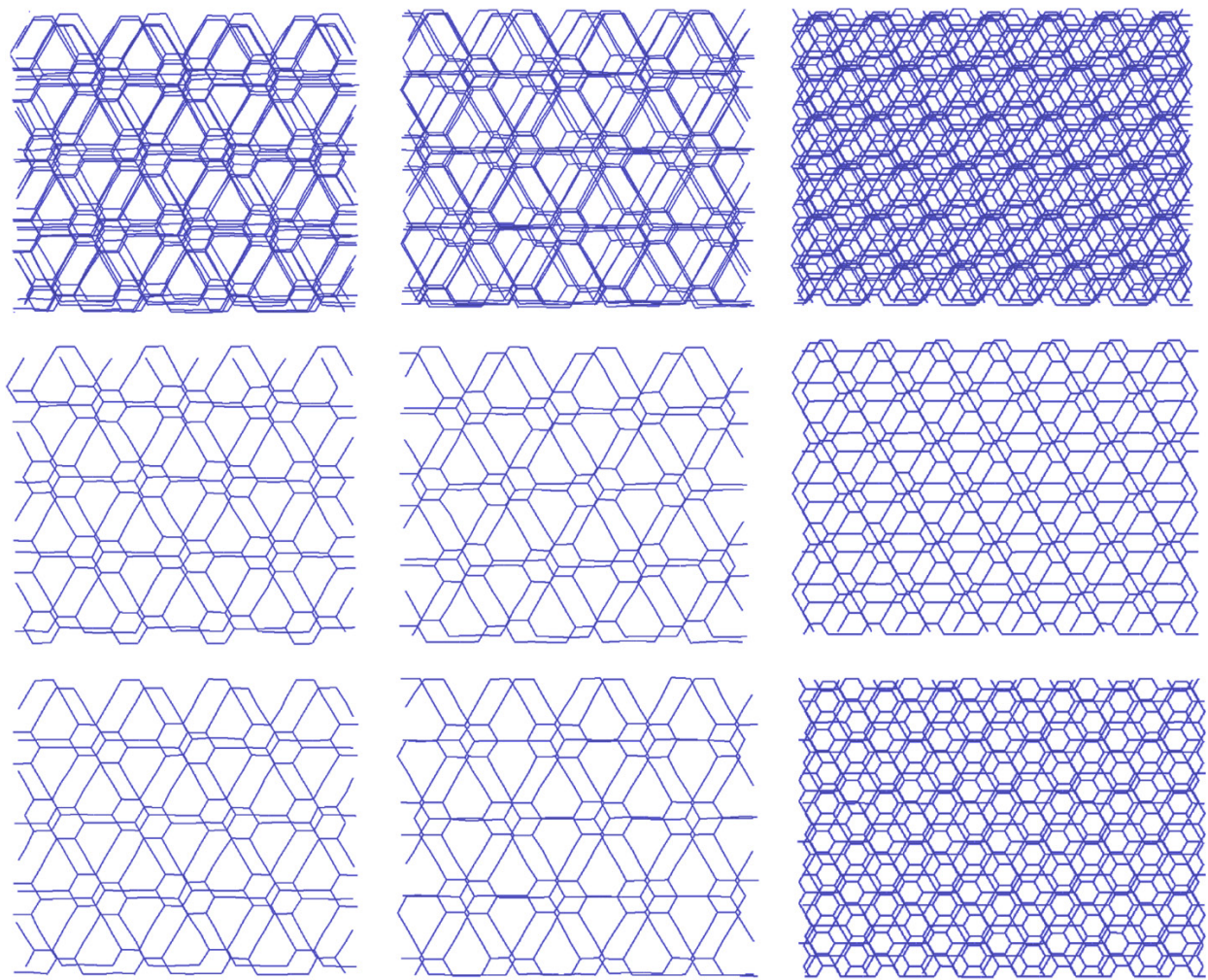

**Fig. S35.** Final frames of MD quench simulations of AA (left), AB1 (middle) and ABC (right) stacking structures of γ-graphynes with 6 layers. The top row shows all 6 layers. The middle and bottom rows show pairs (or triples in the bottommost right structure) of layers to help visualize the local mode of stacking.

Discussion of MD Simulation Results

Table **S10** shows the cohesion energies of the initial stacking structures (referred to as AA, AB1, AB2, and ABC, Fig. S34) computed using the ReaxFF potential. ReaxFF predicts that the lowest energy stacking mode is AB2. However, the differences in the formation energies are so small that the structures might be considered energetically equivalent. Furthermore, it is likely that these stacking modes are easily interconvertible.

**Table S10**. ReaxFF energies for possible stacking modes of γ-graphyne.

| Structure | Energy [eV/atom] |
| --- | --- |
| AA | -8.0010 |
| AB1 | -8.0018 |
| ABC | -8.0026 |
| AB2 | -8.0052 |

The dynamics of these stacking structures is interesting. Thermal equilibration or quenching simulations starting from the AA, AB1 or ABC stacking structures end with structures where pairs or triples of layers adopt one of the two smallest energy stackings, AB2 or ABC. Figure S35 shows that under thermal fluctuations, the layers are relatively free to move and find out the best local stacking. Panels in Fig. S35 confirm that the AB2 is the preferable stacking for γ-graphyne bilayer, which agrees with DFT calculation results. In this stacking mode, the upper layer is bound at an A-type site (Fig. 3e, Main Text).